\documentclass[11pt]{article}
\usepackage{amsfonts,amssymb,amsmath,graphics,mathtools}
\usepackage{geometry,graphicx, color, psfrag}
\usepackage{tikz, tkz-euclide}
\usetikzlibrary{calc}
\usetikzlibrary{decorations.markings}

\usepackage[english]{babel}
\usepackage{color}

\usepackage{harpoon}
\newcommand*{\nicevec}[1]{\overrightharp{\ensuremath{#1}}}

\newcommand{\ew}{\color{black}}

\newtheorem{theorem}{Theorem}
\newtheorem{definition}{Definition}
\newtheorem{remark}{Remark}
\newtheorem{proof}{Proof}

\newtheorem{proposition}{Proposition}
\newtheorem{lemma}{Lemma}

\newcommand{\beq}{\begin{eqnarray}}
\newcommand{\eeq}{\end{eqnarray}}

\newcommand{\beqt}{\begin{eqnarray*}}
\newcommand{\eeqt}{\end{eqnarray*}}

\newcommand{\be}{\begin{equation}}
\newcommand{\ee}{\end{equation}}

\newcommand{\bl}{\begin{lemma}}
\newcommand{\el}{\end{lemma}}

\newcommand{\bcon}{\begin{conjecture}}
\newcommand{\econ}{\end{conjecture}}

\newcommand{\br}{\begin{remark}}
\newcommand{\er}{\end{remark}}

\newcommand{\bt}{\begin{theorem}}
\newcommand{\et}{\end{theorem}}

\newcommand{\bd}{\begin{definition}}
\newcommand{\ed}{\end{definition}}

\newcommand{\bp}{\begin{proposition}}
\newcommand{\ep}{\end{proposition}}

\newcommand{\bc}{\begin{corollary}}
\newcommand{\ec}{\end{corollary}}

\newcommand{\bpr}{\begin{proof}}
\newcommand{\epr}{\end{proof}}

\newcommand{\bi}{\begin{itemize}}
\newcommand{\ei}{\end{itemize}}

\newcommand{\ben}{\begin{enumerate}}
\newcommand{\een}{\end{enumerate}}

\newcommand{\Z}{\mathbb Z}
\newcommand{\R}{\mathbb R}
\newcommand{\N}{\mathbb N}

\newcommand{\s}{\ensuremath{\mathcal{S}}}

\newcommand{\om}{\ensuremath{\omega}}
\newcommand{\Om}{\ensuremath{\Omega}}

\newcommand{\La}{\ensuremath{\Lambda}}

\begin{document}

\title{{\bf  Decimations for Two-dimensional  Ising and Rotator Models I}}
 
 \author { Matteo D'Achille
 \ew \footnote{LAMA UPEC $\&$ CNRS, Universit\'e Paris-Est,  94010 Cr\'eteil, France,
 \newline
 email:  matteo.dachille@u-pec.fr}~, Aernout C.D. van Enter
\ew\footnote{Bernoulli institute, University of Groningen, 9747AG, Groningen, Netherlands,
\newline
email: avanenter@gmail.com}~, Arnaud Le Ny
 \ew \footnote{LAMA UPEC $\&$ CNRS, Universit\'e Paris-Est,  94010 Cr\'eteil, France,
 \newline
 email:  arnaud.le-ny@u-pec.fr}
}

\maketitle
{\bf Abstract:} We extend proofs of non-Gibbsianness of decimated Gibbs measures at low temperatures to include long-range, as well as  vector-spin interactions.
{Our main tools consist in a two-dimensional use of ``Equivalence of boundary conditions'' in the long-range case and an extension of Global specifications for two-dimensional vector spins.}

  \tableofcontents


\normalsize


\section{Introduction}

\subsection{Questions on Preservation of Gibbsianness under Decimation;\\ Background on  related Issues and Strategy}

{ In this paper, we extend the study of Gibbsian and in particular non-Gibbsian properties of decimated measures. Decimation transformations form an example of Renormalisation Group transformations. Such  Renormalisation Group transformations, widely considered in the physics literature, are supposed  to be well-defined on interactions or interaction parameters.   Although this well-definedness is fairly immediate for hierarchical models \cite{BlSi73,JL75}, mathematically,  on ordinary lattices for most of the transformations used in practice-  for the study of critical behaviour-  this existence issue is not obvious.  Decimation transformations  form  one of the cleanest examples to illustrate these points on. The issue  was first raised and analysed by  Griffiths and Pearce and by Israel, and later systematically studied, and interpreted as the possibility of transformed measures to be {\em non-Gibbsian} in~\cite{EFS93}. The Griffiths-Pearce peculiarities (pathologies) were first presented in  \cite{GP78, GP79}, for  Israel's analysis and example,  see  \cite{Isr81}.}

As was shown there, at high temperature or in strong external fields the transformations are well-defined;  it turned out later that  even in some cases around the critical points decimated Ising and rotator models tend to be Gibbsian. 
For those results see in particular \cite{HallerKen1996,Ken}.
Only for Potts models, and similarly for very nonlinear rotator models, it is known or expected that non-Gibbsianness occurs for decimated measures around the transition temperature.
See \cite{EFK,ES2005}.

In this paper, however, we are mainly considering the low-temperature regimes. 
We note that analogous Gibbs-non-Gibbs questions have been considered for Gibbs measures which evolved under stochastic  evolutions,  like Glauber or interaction diffusion dynamics,  or for coarse-graining maps; this  included some results for vector spins. Also more recently, one-dimensional long-range Ising models were considered.  For some of those results, see 
\cite{ER2008,ER2009,CR2016, 
EKO2011,  KO,
 EKOR2010, ELN17,Ken2020}.

Here we extend these results; we  will consider decimation on even sublattices, acting on two-dimensional  long-range models, for  both Ising and vector spins, for which the strategy of the proofs for the Ising model can be straightforwardly generalised.  In particular, the alternating configuration of decimated (visible) sites  will be a point of discontinuity for the conditional probabilities of the decimated measure --a bad configuration --,  because conditioned on it, the model on invisible sites is equal to the original model  on a periodically dilute, or ``decorated''  lattice,  and has a phase transition at low enough temperatures.\\

As  a tool to be used,  we will extend to the rotator-spin context the concept of {\em Global Specifications}, which were shown to exist for Ising ferromagnets in 
\cite{FP97}.
In a forthcoming Part II, we will consider borderline cases, where the Mermin-Wagner theorem prevents symmetry-breaking in the original model, but nonetheless non-Gibbsian behaviour applies, more similarly to what happens in stochastically evolved measures \cite{ER2008,ER2009,EKOR2010}. \\  In those models, the alternating configuration will not lead to a phase transition of the conditioned model, however, and the search for  ``bad'' configurations is somewhat more involved.  Also the conditioned models will have phase transitions of different types ({\em spin-flop}); this occurrence of different types of transitions is also  similar to what happens in the mean-field context for stochastic evolutions of Curie-Weiss Ising model \cite{KLN2007}.


{Following the general approach of \cite{EFS93} (hereafter ``the EFS approach''), one wants to show that the conditional expectation of some image microscopic variable (spin) at a fixed site (which may be taken to be the origin) is essentially discontinuous as a function of the boundary condition. In the EFS approach one goes through the following steps:  
\begin{enumerate}
\item[\bf Step 0:] One divides spins into ``visible" ones (or renormalised, or evolved, or simply primed) and ``invisible'' ones (to be integrated out, initial, or non-primed). Then one considers the marginal measure on the visible spins (i.e.\ the renormalised, evolved, or primed measure). In this paper we shall consider the decimation transformation and consider the spins on a periodic sublattice. Non-Gibbsianness is thus obtained {\em via} a Conditioned Phase Transition; ({\em invisible} Long-Range Order shows up as a {\em visible} Nonlocality). 

\item[\bf Step 1:] Conditioning events on infinite subgraphs  is not immediate, but it may be  allowed either by the  existence of a Global Specification, or by  the existence  of a well-defined conditoning procedure which can be checked by hand.  Global Specifications were originally introduced  as part of the  results of Fern\'andez-Pfister for monotone attractive specifications {for Ising spins} (see~\cite{FP97}, \S 3.1). In this paper we shall pursue this latter path and extend the approach of Fern\'andez-Pfister to vector spins.  

\item[\bf Step 2:] Phase transition.

Conditioned on some ``bad'' visible configuration there should be a phase transition (that is, different coexisting Gibbs measures) for the invisible spins (phase transitions can be of different types, original spin-flip, spin-flop, etc.), and this is model-dependent. Notably, such a transition can also happen when the original model has no transition. In this paper we shall consider various models with long-range interactions which extends ideas from analysis of earlier $1d$ long-range and $2d$-$n.n.$ models~\cite{EFS93,ELN17}. In a part II companion paper we will analyse cases where the situation about the  phase transition of the conditioned system will be different
 compared to the unconditioned one, 
 as  also occurred in  \cite{ER2008}.

\item[\bf Remark 2a:] Positive results are available about Gibbsianness of transformed measures, {\em via} Absence of Phase Transitions. For example strong absence of phase transitions for all visible configurations implies Gibbsianness, see {\em e.g.}\ Kennedy  {\em et al.}~\cite{HallerKen1996, Ken2020}, Olivieri {\em et al.} \cite{BCO06}. More recently such derivations were extended, and  Gibbsian properties were derived also under weaker conditions, by  Berghout and Verbitskiy \cite{Berghout2020, BV}.

\item[\bf Step 3:] Selection.

Visible-spin configurations in an  annulus around the volume under consideration should be ``good'' and be able to select an invisible unique phase in the annulus, which then acts similarly to pure boundary conditions.  Conditioned on such pure-phase-like visible spins everywhere we have infinite-volume uniqueness of Gibbs measure on invisible spins. This is usually direct on lattices, but can be problematic {\em e.g.}\ on trees. (see {\em e.g.}~\cite{vEEIK2012}), or --as also will be a problem in the next item-- when the visible and invisible spins are decoupled.  Uniqueness can follow either from a Lee-Yang argument, or from a contour argument along the lines of Pirogov-Sinai theory. 

\item[\bf Step 3a:] Extra step in long-range models, leading to a need for growing annuli:\\
A wide annulus is needed for  screening effects.  \\ The annulus should be chosen wide enough, such that on the one hand the {\em direct} interaction between the volume inside and the region outside the annulus is (uniformly) small. This argument is similar to Equivalence of Boundary Conditions.\\On the other hand, the uniqueness of the invisible Gibbs state in the annulus shields the {\em indirect} influence  which can be transmitted from the region outside the annulus to the volume inside {\em via} the invisible spins in the annulus.  \\This second requirement may or may not require a wide annulus in short-range models. The first requirement for a wide annulus is specific,  and always needed, for long-range models. 

\item[\bf Step 4:] Unfix the origin. 

Then the choice of the invisible phase, conditioned on all other visible spins  influences expectation of visible spin at origin.
Usually this poses no problem, but can be problematic, if,  for example,  visible (primed) and invisible (non-primed) spins are not coupled  -independent-.
\end{enumerate}

}

 \subsection{Summary of  Results}


{{In this paper we give several non-Gibbs results at low temperatures, as follows (by increasing order of complexity/peculiarity) :}}
\begin{itemize}

\item {\bf Decimation of the $2d$ long-range Ising model}

In this context, we extend the results of non-Gibbsianness at low temperatures, previously known in $2d$ (and higher $d$) for $n.n.$ models and in $1d$ for long-range models, to various long-range models. In addition to the standard procedure recalled in the introduction, this amounts to control the long-range effects by energy estimates, adapted to each model, in order to also use an argument to control the direct influence from afar,  similarly to the  equivalence of boundary conditions concept coined by Bricmont {\em et al.} and then conclude as in the simpler $2d$-$n.n.$ case. 

The long-range Ising models for which we prove it here are the following:  

{\bf - Bi-axial $n.n.$-long-range Ising models ($\alpha_1>1$) :}

 {\em Bad configuration :} alternating $\omega'_{{\rm alt}}=(-1)^{i_1+i_2}$ for any site $i=(i_1,i_2)$
 
  {\em Ferromagnetic couplings considered :} ($d=2$, $\alpha_1 > 1$) : $J \geq 0$,
 $$
 J^{n.n.,\alpha_1}(i,j)\coloneqq J \cdot  \mathbf{1}_{|i -j |=1} \cdot \mathbf{1}_{|i_1-j_1|=0}+ J \cdot |i_1-j_1|^{-\alpha_1}  \cdot \mathbf{1}_{|i_2-j_2|=0} 
 $$

 We get non-Gibbsianness at low enough temperatures for all  $\alpha_1 > 1$.
 
{\bf - Axially long-range Ising models (possibly anisotropic $\alpha_1 \neq \alpha_2$) :}

  {\em Bad configuration :} alternating $\omega'_{{\rm alt}}=(-1)^{i_1+i_2}$ for any site $i=(i_1,i_2)$, 
  
  {\em Ferromagnetic couplings considered :} ($d=2$, $\alpha_1,\alpha_2 > 1$) : $J \geq 0$,
 $$
 J^{\alpha_1,\alpha_2}(i,j)\coloneqq J \cdot  |i_2-j_2|^{-\alpha_2}  \cdot \mathbf{1}_{|i_1-j_1|=0}+ J \cdot |i_1-j_1|^{-\alpha_1}  \cdot \mathbf{1}_{|i_2-j_2|=0}
 $$

  We get non-Gibbsianness at low enough temperatures or all $\alpha_1,\alpha_2>1$  
 
{\bf - Isotropic long-range Ising models :}

  {\em Bad configuration :}  alternating $\omega'_{{\rm alt}}=(-1)^{i_1+i_2}$ for any site $i=(i_1,i_2)$
 
  {\em Ferromagnetic couplings considered :} ($d=2$, $\alpha >2$) : $J \geq 0$,
 {$$
 J^{iso,\alpha}(i,j)\coloneqq    J \cdot |i_1-j_1|^{-\alpha}  \cdot \mathbf{1}_{|i_2-j_2|=0} 
 $$}
 We get non-Gibbsianness at low enough temperatures for all $\alpha > 2$.

 \item {\bf Decimation of   anisotropic/long-range rotator models with phase transitions} 
 
 Then we turn to the extension of non-Gibbsianness in the case of $2d$ rotator models. The crucial advantage of the fact that we consider two-dimensional vector spins
 in our ferromagnetic context is the stochastic ordering which occurs, 
 which will allow a proper extension of the local specification into a global one ({see next section}). 
 
 {\bf - $2d$-$n.n.$ Anisotropic rotator models}
 
{ Here  we know that a phase transition with only two extremal measures occurs, with some possible stochastic ordering, and this allows to extend the existing  $2d$-$n.n.$  proof \cite{EFS93} {using global specifications as in} \cite{ALN13}. This original proof does not require energy estimates {or} equivalence of b.c., as the specification is Markov.}

 {\em Bad configuration :} alternating $\omega'_{{\rm alt}}$ with angle $\theta_i=(-1)^{i_1+i_2} \frac{\pi}{2}$ for any site $i=(i_1,i_2)$

{\em Ferromagnetic Pair potential :} $\Phi_A=0$ unless $ A=\{i,j\}, |i-j|=1$, {$\kappa \in (0,1)$} and
 $$
 \Phi_{\{i,j\}}(\nicevec{\sigma})= - J \langle \nicevec{\sigma}_{i} \cdot \nicevec{\sigma}_{j} \rangle_{\kappa} =- J \big( \sigma_{i_1}\sigma_{j_1}+\kappa\, \sigma_{i_2}\sigma_{j_2} \big).
$$ 
 
We get non-Gibbsianness at low enough temperatures {for any $\kappa \in (0,1)$}.
 
{\bf - $2d$ planar long-range rotator models}

There, we use the spontaneous magnetization at low temperature derived by Kunz and Pfister \cite{KP76} by  comparison with a hierarchical model, {\em \`a la} Dyson, coupled to the global specification. For both ingredients, we emphasized again that the correlation inequalities for vector spins derived by Ginibre, and not valid at higher dimensions, are essential.

 {\em Bad configuration :} alternating $\omega'_{{\rm alt}}$ with angle $\theta=(-1)^{i_1+i_2} \frac{\pi}{2}$ for any site $i=(i_1,i_2)$
 
{\em Ferromagnetic couplings :} ($\alpha > d=2$) : $J \geq 0$, $ i,j \in \mathbb{Z}^d$,
 $$
J^{iso,\alpha}(i,j)\coloneqq  J \cdot |i-j|^{-\alpha} \cdot \langle \nicevec{\sigma}_{i} \cdot \nicevec{\sigma}_{j} \rangle.
 $$
  We get non-Gibbsianness at low enough temperatures for all $\alpha \in (2,4)$.

 \end{itemize}

\section{General Framework and Global Specifications}

\subsection{General Framework}

We will focus on  Ising and vector spins on the $d$-dimensional lattice $\Z^d$, mostly concentrating on the planar case ($d=2$). As usual in Mathematical Statistical Mechanics, we investigate   infinite-volume behaviors, and in particular consider  the {\em Dobrushin-Lanford-Ruelle} (DLR) framework  \cite{Dob68,LR69}, where at infinite volume, Gibbs measures  are defined by means of the specification of the  regular versions of their conditional probabilities w.r.t.\ the outside of finite sets (on which boundary conditions are prescribed). 

{\bf {\em Lattice Structure:}}

We denote by $\s$ the set consisting of  all finite subsets of the lattice $\Z^d$, and we will consider often sequences of cubes $\Lambda_L=([-L,+L] \cap \Z)^d, \; L \in \N$, to perform {\em Thermodynamic Limits}\footnote{In this limiting procedure, one has to   respect convergence to zero of the ratio  surface/volume. Sometimes less stringent convergence along a net directed by inclusion can be enough, see \cite{EFS93}.}. {Lattice} sites will be denoted by {Latin} letters $i,j,k$,   with components $i_j=(i_1,\dots,i_d)$ (mostly $i=(i_1,i_2)$). We denote by $|\cdot|$ the $\mathbb{L}^2$-norm on $\Z^d$ . When $|i-j|=1$, the  two sites   $i,j \; \in \Z^d$
are said to be {\em nearest neighbors} (briefly written $n.n.$).

{\bf {\em Measurable Structure :}}

We shall consider two different single-site state spaces $(E,\mathcal{E}, \rho_0)$, modelling two different sort of microscopic values (``spins''): 

{\bf Ising spins :} The state space is the simple alphabet $E=\{-1,+1 \}$, equipped with the {\em a priori} counting measure $\rho_{0}=\frac{1}{2}\delta_{-1}+\frac{1}{2}\delta_{+1} $ and $\mathcal{E}=\mathcal{P}(\{-1,+1\})$.

or  

{\bf Rotator vector --$XY$-- spins :}  The state space is the unit circle $E=\mathbb{S}^{1}$, equipped with the Borel $\sigma$-algebra $\mathcal{E}$ and with the normalized Haar measure as  {\it a priori} measure $\rho_0$. 

In this vectorial case, we denote by $(\nicevec{e_1}, \nicevec{e_2})$ the canonical basis of $\mathbb{R}^2$ and  to pursue the analogy with the Ising case, we sometimes identify the sphere $E$ with $]-\pi,+\pi]$. In this case, for $i \in \Z^2$, we shall identify a spin vector $\nicevec{\sigma}_i $ in $E$ by its angle $\theta_i$ with the horizontal element of the basis, 
$$
\theta_i = \theta(\nicevec{\sigma}_i) = (\nicevec{\sigma}_i,\nicevec{e_1}) \in ]-\pi,+\pi]
$$
where $(\cdot , \cdot )$ will denote the angle between vectors (not to be confused with scalar products $\langle \cdot,\cdot \rangle$ used below in the interactions).

To each site $i$ of the lattice is attached a spin with values $\sigma_i \in E$ (resp.\ $\nicevec{\sigma}_i \in E$), giving rise to infinite-volume configurations of the form $\sigma=(\sigma_i)_{i \in \Z^d}$ (resp.\ $\nicevec{\sigma}=(\nicevec{\sigma}_i)_{i \in \Z^d} $).  Microscopic states will sit in  the (infinite-volume) configuration space which will be  the infinite-product probability space equipped with the product measurable structure, 
$$
(\Omega,\mathcal{F},\rho) = (E^{\Z^d},\mathcal{E}^{\otimes \Z^d}, \rho^{\otimes \Z^d}).
$$

These configurations are denoted generically by Greek letters $\sigma,\omega$, etc., or  $\nicevec{\sigma}, \nicevec{\omega}$, etc. in the rotator cases. They are infinite families of {\bf random variables} $ {\sigma}= ({\sigma}_i )_i \in \{-1, +1\}^{\Z^d}$ in the Ising case or {\bf random vectors} ${\bf \nicevec{\sigma}}= (\nicevec{\sigma}_i )_i \in (\mathbb{S}^1)^{\Z^d}$ in the rotator case. We also denote $(\Omega_\Lambda, \mathcal{F}_\Lambda, \rho_\Lambda)=(E^\Lambda, \mathcal{E}^{\otimes \Lambda}, \rho^{\otimes \Lambda})$ to be the restriction/projection of $\Omega$ on $\Omega_\Lambda$, for $\Lambda \in \s$.  We also generically consider possibly infinite subsets $\Delta \subset \Z^d$, for which all the preceding notations defined for finite $\Lambda$ extend naturally ($\Omega_\Delta,\mathcal{F}_\Delta, \rho_S,\sigma_\Delta$, etc.).

 On these single-spin state spaces $E$, we shall consider pair potentials with different types of couplings $J(i,j)$ (isotropic {\em vs.}\ anisotropic, or $n.n.$ {\em vs.} long-range --polynomially decaying), but always in a ferromagnetic context (with some positive measurable functions $J$ on $\Z^d \times \Z^d$, with $J(i,j) \geq0$ for any pair $\{i,j\}$, see next sections).

We also denote by $\mathcal{M}_1^+$ 
 the set of   probability measures on $(\Omega,\mathcal{F})$. We moreover consider  some partial order $\leq$ on $\Omega$ : $\sigma \leq \omega$ if and only if $\sigma_i \leq \omega_i$ for all $i \in \Z^2$. This order extends to functions: $f$ is called {\em increasing}  when $\sigma \leq \omega$ implies $f(\sigma) \leq f(\omega)$. It induces then a {\em stochastic order} on measures and we write $\mu \leq \nu$ if and only if it is valid for expectations, with  $\mu[f] \leq \nu[f]$ for all $f$  increasing\footnote{{We only need to consider real-valued functions here --even in the rotator case where we consider vertical magnetisations-- but the extension to vector-valued functions is straightforward componentwise.}}. The Gibbsian formalism we consider here, built within the DLR framework, has been fully  rigorously described by Georgii \cite{Geo88}; macroscopic states are modelled by Gibbs measures inspired by a mix of measurable and topological considerations (see also \cite{Fer05}), with the important property of continuity for regular versions of finite-volume conditional probabilities as a function of the boundary condition, in the sense more precisely given below.

{\bf {\em Topological Structure :}}

Our configuration spaces will be endowed with the product topology of the canonical topology of the underlying single-spin state space $E$, {\em i.e.} the discrete topology in the Ising cases, and the Borel topology on the circle {$\mathbb{S}^{1}$} in the the rotator cases. 

In this  product topology {of the discrete topology on $E=\{-1,+1\}$ (resp.\ the Borel topology on $E=]-\pi,+\pi]$)}, configurations are close when they coincide ({resp.\ when they are close}) on large finite regions $\Lambda$ and arbitrary outside. Of course, the larger the finite common  region, the closer they are. We denote by $C(\Om)$ the set of continuous  functions on $\Om$ equipped with these topologies.

{\bf In the Ising context}  with finite state-space equipped with the discrete topology, continuity is equivalent to uniform continuity and to so-called {\em quasilocality}, defined as

\be \label{qlocfu}
f \in C(\Omega) \; \Longleftrightarrow \; \lim_{\Lambda \uparrow \s} \sup_{\sigma,\omega:\sigma_\Lambda=\omega_\Lambda} \mid f(\omega) - f(\sigma) \mid = 0.
\ee
Quasilocality itself is closely related to the concept of Gibbs measures, as we shall see. 

 For a given configuration $\omega \in \Om$, a  neighbourhood basis  is provided by the family\\ $\big(\mathcal{N}_\Lambda(\omega)\big)_{\Lambda \in \mathcal{S}}$ with, for any $\Lambda \in \s$,
$$
\mathcal{N}_\Lambda(\omega)=\Big \{ \sigma \in \Omega : \sigma_\Lambda=\omega_{\Lambda},\; \sigma_{\Lambda^c} \; {\rm arbitrary} \Big\}.
$$
We also consider {in this case} particular open subsets { $\mathcal{N}_{\Lambda,\Delta}^+(\omega), \mathcal{N}_{\Lambda,\Delta}^-(\omega)$} of {the} neighborhoods $\mathcal{N}_\Lambda(\omega)$ on which {configurations also coincide with the maximal $+$-configuration (resp.\ $-$-configuration}) on an annulus $\Delta \setminus \Lambda$ for $\Delta \supset \Lambda$, defined for all $\Lambda \in \s,\; \om \in \Om$ as

\begin{eqnarray*}
\mathcal{N}_{\Lambda,\Delta}^+(\omega)&=&\Big \{ \sigma \in \mathcal{N}_\Lambda(\omega) : \sigma_{\Delta \setminus \Lambda } = +_{\Delta \setminus \Lambda},\; \sigma_{\Delta^c} \; {\rm arbitrary} \Big\},\\
\mathcal{N}_{\Lambda,\Delta}^-(\omega)&=&\Big \{ \sigma \in\mathcal{N}_\Lambda(\omega) : \sigma_{\Delta \setminus \Lambda } = -_{\Delta \setminus \Lambda},\; \sigma_{\Delta^c} \; {\rm arbitrary} \Big\}.
\end{eqnarray*}

 {\bf In the rotator context} for the Borel topology  on the sphere {$\mathbb{S}^1$} or on the interval $]-\pi,+\pi]$, continuity is stronger than quasilocality, for which the  definition  (\ref{qlocfu}) still holds. In both contexts, the set of measurable quasilocal functions, denoted $\mathcal{F}_{{\rm qloc}}$, is also the set of the possible uniform limits of local functions, the one that are $\mathcal{F}_\Lambda$-measurable for some $\Lambda \in \s$. Note that in the  rotator context there exist local and quasilocal functions which are not continuous, such as characteristic functions of the spin at the origin being inside some interval, see \cite{Geo88, EFS93}.

For a given configuration $\nicevec{\omega} \in \Om$, to get  ``open subsets'' with positive measure, one cannot ask anymore that angles are fixed, so they are only constrained in small intervals around the configuration. A  basis of neighborhoods is {then} provided for  a given parameter sequence  $\epsilon_k >0$   by the family  $\big(\mathcal{N}_{\Lambda,\epsilon_k}(\nicevec{\omega})\big)_{\Lambda \in \mathcal{S}}$ with, for any $\Lambda \in \s$,
$$
\mathcal{N}_{\Lambda,\epsilon_k}(\nicevec{\omega})=\Big \{ \nicevec{\sigma}  \in \Omega :  (\nicevec{\sigma}_i,\nicevec{\omega}_i)\leq \epsilon_k, \forall i \in \Lambda;    \nicevec{\sigma}_{\Lambda^c} \; {\rm arbitrary} \Big\}.
$$

{

We also consider particular open subsets of neighborhoods $\mathcal{N}_{\Lambda,\epsilon} (\nicevec{\omega})$ for which, on the contrary to the Ising case,
the angles are not fixed on the annulus  but rather confined in small intervals of radius $\epsilon >0$ around the maximal and minimal values, that is configurations are also close to specific  configurations of canonical angles $\pm \pi/2$ on an annulus $\Delta \setminus \Lambda$ for $\Delta \supset \Lambda$, defined for all $\Lambda \in \s,\; \om \in \Om$ as
\begin{eqnarray*}
\mathcal{N}_{\Lambda,\Delta,\epsilon}^{+\frac{\pi}{2}}(\nicevec{\omega})\coloneqq \Big \{ \nicevec{\sigma} \in \mathcal{N}_{\Lambda,\epsilon}(\nicevec{\omega}) : (\nicevec{\sigma}_i, \nicevec{e_1}) \in \Big(+\frac{\pi}{2}-\epsilon,+\frac{\pi}{2}+\epsilon\Big) \; {\rm for} \; i \in {\Delta \setminus \Lambda },\; \nicevec{\sigma}_{\Delta^c} \; {\rm arbitrary} \Big\},\\
\mathcal{N}_{\Lambda,\Delta,\epsilon}^{-\frac{\pi}{2}}(\nicevec{\omega})\coloneqq \Big \{ \nicevec{\sigma} \in \mathcal{N}_{\Lambda,\epsilon}(\nicevec{\omega}) : (\nicevec{\sigma}_i, \nicevec{e_1})\in \Big(-\frac{\pi}{2}-\epsilon,-\frac{\pi}{2}+\epsilon\Big) \; {\rm for} \; i \in {\Delta \setminus \Lambda },\; \nicevec{\sigma}_{\Delta^c} \; {\rm arbitrary} \Big\}.
\end{eqnarray*}

We shall sometimes shortly denote $\mathcal{N}_{\Lambda,\Delta,\epsilon}^+(\nicevec{\omega})\coloneqq \mathcal{N}_{\Lambda,\Delta,\epsilon}^{+\frac{\pi}{2}}(\nicevec{\omega})$ or $\mathcal{N}_{\Lambda,\Delta,\epsilon}^-(\nicevec{\omega})\coloneqq \mathcal{N}_{\Lambda,\Delta,\epsilon}^{-\frac{\pi}{2}}(\nicevec{\omega})$.


{\bf {\em Macroscopic States :}}
  
  In Mathematical Statistical Mechanics, macroscopic states are thus represented by  measures in  $\mathcal{M}_1^+$. To describe such measures on the infinite-product probability space $\Omega$, in view of a  mathematical description of phase transitions and phase coexistence, one aims  at describing it by prescribing versions of conditional probabilities w.r.t.\  boundary conditions outside finite sets. In this so-called {\em DLR approach}, independently introduced  in the late 60's  by Dobrushin \cite{Dob68} in the East, and Lanford/Ruelle \cite{LR69} in the West, candidates to represent such a system of conditional probabilities are families of probability kernels, formally introduced by F\"ollmer \cite{Foll75} and Preston \cite{Pres76} in the mid 70's under the terminology 
 {\em (local) specifications} :
\begin{definition}[(Local) Specification]
A {\em specification} $\gamma=\big(\gamma_\Lambda\big)_{\Lambda \in \s}$  on $(\Omega,\mathcal{F})$ is a family of probability kernels  $\gamma_\Lambda : \Omega \times \mathcal{F} \; \longrightarrow \; [0,1];\; (\omega,A) \; \longmapsto \;=\gamma_\Lambda(A \mid \omega)$
s.t. for all $\Lambda \in \mathcal{S}$:
\begin{enumerate}
\item For all $\omega \in \Omega$, $\gamma_\Lambda(\cdot | \omega)$ is a probability measure on $(\Omega,\mathcal{F})$.
\item For all $A \in \mathcal{F}$, $\gamma_\Lambda(A | \cdot)$ is $\mathcal{F}_{\Lambda^c}$-measurable.
\item {(Properness)} For all $\omega \in \Omega$, $\gamma_\Lambda(B|\omega)=\mathbf{1}_B(\omega)$ when $B \in \mathcal{F}_{\Lambda^c}$.
\item { (Consistency)} For all $\Lambda \subset \Lambda' \in \s$, $\gamma_{\Lambda'} \gamma_{\Lambda}=\gamma_{\Lambda'}$, where 
\be \label{DLR0}
\forall A \in \mathcal{F},\; \forall \omega \in \Omega,\;(\gamma_{\Lambda'} \gamma_\Lambda)(A | \omega)=\int_\Omega \gamma_\Lambda(A | \omega') \gamma_{\Lambda'}(d \omega' | \omega).
\ee
\end{enumerate}
\end{definition}
These kernels also acts on functions and on measures: for all $f \in C(\Omega)$ or $\mu \in \mathcal{M}_1^+$,
$$
\gamma_\Lambda f(\omega)\coloneqq \int_\Omega f(\sigma) \gamma_\Lambda (d \sigma | \omega)=\gamma_\Lambda [f | \omega] \; {\rm and} \; 
\mu \gamma_\Lambda [f] \coloneqq \int_\Omega (\gamma_\Lambda f)(\omega) d \mu (\omega)= \int_\Omega \gamma_\Lambda [f | \omega] \mu(d \omega).
$$

{ Following Fern\'andez-Pfister \cite{FP97}}, a local specification is said to be {\em monotonicity-preserving} or {\em attractive}\footnote{{See also the books by Preston \cite{Pres75,Pres76}.}} if, for all $\Lambda \in \s$ and $f$ increasing, the function $\omega \mapsto  \gamma_\Lambda f(\omega)$ is an increasing function of the boundary condition $\omega$. It is straightforward that Gibbsian specifications for the  ferromagnetic pair-potentials ({\em i.e.\ }with coupling functions $J(i,j)  \geq 0$) considered here are monotonicity-preserving.

In order to extend local specifications to global ones in these contexts, and to be able to  profit from monotone convergence theorems, extending \cite{FP97} beyond  the Ising-spin case with finite alphabet, we shall need to identify an underlying partial order on configurations,  especially in the rotator case (where it is not so obvious, and restricted to spin dimension two)}, see Theorem \ref{globspe-rotator}.

{\bf In the Ising case} one can take the canonical order $\leq$ on $E$ and says that $\omega \leq \omega'$ if and only if $\omega_{i} \leq \omega'_{i}$, for all $i \in \Z^2$. The two ``extremal'' (i.e.\ minimal and maximal) configurations, denoted ${\bf -}$ and ${\bf +}$ and defined as ${\bf -}_i$=-1 for all $i \in \Z^2$ and ${\bf +}_i=+1$ for all $i \in \Z^2$, will give rise to infinite-volume limits $\mu^-$ and $\mu^+$ that are extremal in two ways :  first, they are extremal Gibbs measures; and second, they are extremal with respect to this partial order as, for any other Gibbs measure (describe below) $\mu$,  it holds that  $\mu^-\leq \mu \leq \mu^+$, {\em i.e.} for all $f$ increasing, 
$$
\mu^-[f] \leq \mu[f] \leq \mu^+[f].
$$

For a proof that ferromagnetic Gibbs specifications are indeed monotonicity-preserving in this context see {\em e.g.} the original works of Ginibre \cite{Gin71} or Griffiths \cite{Griff68}.

{\bf In the rotator case}, {with state-space $\mathbb{S}^1$}, we need thus to cook up a partial order for which our particular homogeneous configurations ${\bf - \frac{\pi}{2}}$ and ${\bf + \frac{\pi}{2}}$ are the extremal ones with respect to this order. To do so, we choose the partial order $\leq_{\sin}$ as follows : 

Let $\theta=\theta_{i} \in ]-\pi,+\pi]$ be the canonical angle related to the configuration $\omega_{i}$ at site $i$, and let $\theta'_{i}$ be the corresponding value for $\omega'_{i}$.  Then we say that 

$$
\omega \leq_{\sin} \omega' \; {\rm if \;  and \; only \; if} \; \sin \theta_{i} \leq \sin \theta'_{i},\; {\rm for \; all} \;   i \in \Z^2
$$ 
 and similarly for measures, 
  $$\mu \leq_{\sin} \mu'   \; {\rm if \;  and \; only \; if \; for \; any} \;  f \; {\rm increasing}, \mu[f] \leq  \mu'[f].$$
 
 Note that we keep the expression $\leq_{\sin} $ for the stochastic order (on measures), while it does not appear in the order on expectations. For the latter, it corresponds indeed to the standard order between real numbers, but we stress it has to be tested on functions which are increasing {\em w.r.t. the specific $\leq_{\sin} $ order.}

For such a partial order $\leq_{\sin}$, 
the $XY$-specification with ferromagnetic couplings $J=(J(i,j))_{i,j \in \mathbb{Z}^{2}}$, with $J(i,j) \geq 0$ for any pair $(i,j) \in \Z^2$, is monotonicity-preserving. This allows us  to prove the existence as weak limits   of  {our particular infinite-volume measures $\mu^-$ and $\mu^+$ obtained by taking respectively ${\bf - \frac{\pi}{2}}$ and ${\bf + \frac{\pi}{2}}$-boundary conditions. {The limits are known to exist and to be extremal in the original (convex)  sense since \cite{Griff71}, see also \cite{BLP81}.}

Moreover, these weak limits coincide with the minimal and maximal measures in the sense of the stochastic order : Any other Gibbs measure $\mu$ satisfies, for all $f$ increasing  {\bf in the sense of $\leq_{\rm sin}$,}
$$
\mu^-[f] \; \leq \mu[f] \; \leq \mu^+[f]
$$}
so that the inequalities $
\mu^- \leq_{{\rm sin}} \mu \leq_{{\rm sin}} \mu^+
$ hold.
Note again that this construction is only valid for two-component (rotator, or $XY$) vector spins. For discussions on the extension of  correlation inequalities to higher-dimensional-vector cases, {see  \cite{ BM2012, BG72, BF81, BF82, BFL77, Dun76,  DunNew75, Gin70, HK51, KP76, KPV75, KPV76, New75, Pfis81}}.


For a given measure
one can show that it is always possible to build a consistent specification with a given probability measure \cite{Gold78,Pres80,Sok81}. Nevertheless, different measures can then have their conditional probabilities described by the same specification but on different full measure sets, leaving the door open to a mathematical description of phase transitions, as we shall see below for our ferromagnetic Ising models and {\em some} of our rotator models on $\Z^2$.

\begin{definition}[DLR Measures]
A probability measure $\mu$ on $(\Omega,\mathcal{F})$ is said to be {\upshape \bf consistent} with a specification $\gamma$ (or specified by $\gamma$) when for all $A \in \mathcal{F}$ and $\Lambda \in \s$
\be \label{DLR1}
\mu[A|\mathcal{F}_{\Lambda^c}](\omega)=\gamma_\Lambda(A|\omega), \; \mu{\rm -a.e.} \;  \omega \in \Omega.
\ee
Equivalently, $\mu$ is consistent with $\gamma$ if 
$$
\int (\gamma_\Lambda f) d \mu = \int f d \mu \quad
 \mbox{for all} \quad \Lambda \in \mathcal{S} \quad \mbox{and} \quad f \in \mathcal{F}_{\rm{loc}},
 $$ 
 or if and only if $\mu \gamma_\Lambda = \mu, \; \forall \Lambda \in \s$. 
\end{definition}

We denote by $\mathcal{G}(\gamma)$ the set of measures consistent with $\gamma$. Describing this set is precisely the central task of Equilibrium Mathematical Statistical Mechanics. Indeed, in contrast to {Kolmogorov's Extension Theorem} based on marginals, the existence of a measure for a given specification is not guaranteed nor is the uniqueness:  in particular, one can also get more than one element, and in such a case we say that there is a  {\em Phase Transition}. Central in statistical mechanics, this notion is also  essential for many proofs to get non-Gibbsianness, when phase transitions  occur in some hidden, constrained system, as we shall see in all the remaining sections.

Due to this non-uniqueness phenomenon, the structure of $\mathcal{G}(\gamma)$ can be very rich, see {\em e.g.} Chapter 7 of \cite{Geo88}. In particular, $\mathcal{G}(\gamma)$ is a convex set whose extremal elements are the Gibbs measures that are trivial w.r.t.\ the asymptotic $\sigma$-algebra $\mathcal{F}_\infty= \cap_{\Lambda \in \s} \mathcal{F}_{\Lambda^c}$, and interpreted as the effective {\em physical phases} of the system. Note that we sometimes employ the terminology {\em states} for measures. { Describing the set of such extremal states is a vibrant field of research}, as the set of all Gibbs measures for a given interaction can be very huge.

 { As we shall briefly describe below, the full convex structure for the $2d$-$n.n.$ Ising model can be obtained {\em via} the celebrated Aizenman-Higuchi theorem \cite{Aiz,Hig}}: $\mathcal{G}(\gamma)$ coincides with the interval $[\mu^-,\mu^+]$, thus with ``only'' two extremal elements, that are translation-invariant, and no non-translation-invariant states (and thus there does not exist non-translation-invariant Gibbs measures).
 
  Next to the $2d$-$n.n.$ Ising model, also long-range polynomial Ising models in dimension one (sometimes called ``Dyson'' {or  ``Dyson-Ising"} models {\cite{ELN17,BEELN2018}}) share this property, but this is not the case for higher-dimensional  Ising or rotator models. Indeed, there do exist extremal non-translation invariant states for the three- and {higher}- dimensional $n.n$ Ising models (including -- in $d=3$ probably exclusively -- the so-called Dobrushin/interface states), and uncountably many extremal translation-invariant magnetized states for long-range or high-dimensional rotators at low temperature\footnote{Note it is not rigorously known whether  there are states  different from  the interface ones. For the usual $n.n.$  rotator model those won't exist, but high-dimensional vortices might be thermally stable, {\cite{FrPf83}}.}.  Short-range models in high-temperature or high-field/low-density regions, or in $d=1$, have generally uniqueness, so for those models also the sets of all Gibbs measures are known.   One advantage of working with rotators ($XY$-models) over higher-dimensional vector models such as the classical Heisenberg model, is the existence of FKG ordering -or attractivity-, which does not hold in general vector models, see \cite{KP76, Simon2020}. See also \cite{FV16} for a more complete picture. 

The concept of quasilocality naturally extends from the definition for functions, {\em cf.} (\ref{qlocfu}), to specifications and measures, and provides a proper framework to insure existence of DLR-consistent measures, {\em i.e.} that   $\mathcal{G}(\gamma) \neq \emptyset$. A specification is said to be quasilocal  when the set of quasilocal functions is conserved by its kernels. {More formally, for all $\Lambda \in S$, the image of any local function $f$ {\em via} the kernel of $\gamma$ is a quasilocal function\footnote{{ Thus $\gamma_\Lambda f \in C(\Omega)$ in the Ising case (but not necessarily in the vector case). }} (w.r.t.\ to the boundary condition):
\be
\label{qlocmes}
f  \in \mathcal{F}_{\rm{loc}} \implies \gamma_\Lambda f \in \mathcal{F}_{{\rm qloc}}.
\ee
}

{\bf A measure is said to be quasilocal {iff} it is specified by a quasilocal specification.
}

{{\emph{Gibbs measures}} are measures consistent with a Gibbs specification defined in terms of a uniformly absolutely convergent potential $\Phi$, for which one can give sense to the  {\em Hamiltonian at volume $\Lambda \in \s$ with boundary condition $\omega$} defined, for all $\sigma \in \Om$, as 
\be \label{Hambc}
H_\Lambda^\Phi(\sigma | \omega) \coloneqq  \sum_{A \cap \Lambda \neq \emptyset} \Phi_A(\sigma_\Lambda \omega_{\Lambda^c}) (< \infty)
\ee
{ where $\sigma_\Lambda \omega_{\Lambda^c}$ is the configuration agreeing with $\sigma$ on $\Lambda$ and with $\omega$ on $\Lambda^{c}$}.}
{In this paper, we restrict ourselves to pair-potentials $\Phi^J$ with {\em ferromagnetic} coupling functions $J: S \times S \longrightarrow \R_+$ and formal Hamiltonian

$$
H_\Lambda^J(\sigma) = - \sum_{\{i,j\}} J(i,j) \; \langle {\sigma}_i \cdot  {\sigma}_j \rangle
$$

where ``$\langle \ \cdot \ \rangle $" is either {ordinary product of real numbers} (for Ising spins) or {\em an} inner product (for vector spins), as we shall see.} The {\em Gibbs specification at inverse temperature $\beta>0$} is then given by $\gamma^J=\gamma^{\beta \Phi}$, defined {at finite  volume $\Lambda$} by
\be \label{Gibbspe}
\gamma_\Lambda^{J}(d\sigma \mid \omega)=\frac{1}{Z^{\beta \Phi}_\Lambda(\omega)} \; e^{-\beta H_\Lambda^\Phi(\sigma | \omega)} (\rho_\Lambda\otimes \delta_{\omega_{\Lambda^c}}) (d \sigma)
\ee
where the normalization $Z_\Lambda^{\beta \Phi}(\omega)$ --the partition function-- is a normalizing constant (exponentially) related to a free energy.  Such a specification is {\em non-null} ($i.e.$ when for all  $\Lambda \in \s$ and all $ A \in \mathcal{F}_\Lambda$, $\rho(A)>0$ implies that $\gamma_\Lambda (A | \omega) >0$ for any  $\omega \in \Om$) and has the property that it is {\em quasilocal} (see below). In fact, in the mid-seventies,  Kozlov \cite{Ko} and Sullivan \cite{Su} established that, {for a measure $\mu$}, {\bf  being Gibbs is in fact also equivalent to being non-null and quasilocal}, so that one has the 
\begin{definition}[Gibbs Measure] 
$\mu \in \mathcal{M}_1^+$ is a Gibbs measure iff $\mu \in \mathcal{G}(\gamma)$,  where $\gamma$ is a non-null and quasilocal specification.
\end{definition}
Non-nullness   allows  a proper exponential factor to alter the product structure of the measure -- to get correlated random fields --, while quasilocality allows us to interpret Gibbs measures as natural extensions of the class of  Markov fields.  Sullivan used the term of {\em Almost Markovian} instead of quasilocal in \cite{Su} and provided an uniformly convergent potential from such a non-null specification, while Kozlov \cite{Ko} provided an uniformly absolutely convergent telescoping potential that is not in general translation-invariant, unless the specification is more regular than continuous (see \cite{Fer05} for a complete description, or the recent re-visit of these conditions by Barbieri {\em et al.} \cite{BGMMT}).

In the next sections, we shall explicitly describe  our Gibbs measures, focusing on pair potentials, in either an Ising-spin set-up (with values in the elementary alphabet $E=\{-1,+1\}$), { or in} $O(N)$-{ symmetric} set-ups (where microscopic variables take values in the sphere $N$-dimensional unit sphere, $E=\mathbb{S}^{N-1}$), essentially in dimension two and for $N=2$ (\emph{rotator models}). We shall submit these Gibbs measures to the elementary renormalisation transformation, the decimation of spacing 2, and show  that the transformed measures are not necessarily quasilocal. {For these general descriptions, generic configurations will be written by Greek letters, whatever they concern scalar (Ising) or vector (rotator) spins. We shall indicate afterwards when they are scalar or vector, but not always.}

{\bf {\em Essential discontinuity -- Non-Gibbsianness:}}

Assume that a given specification  $\mu \in \mathcal{G}(\gamma)$ is quasilocal, then for any  $f$ local and  $\Lambda \in \s$,  the conditional  expectations of $f$ w.r.t.\ the outside of $\Lambda$ are $\mu$-a.s. given by $\gamma_\Lambda f$,   by the DLR Equations (\ref{DLR1}), and it is itself a  {quasilocal}  function of the boundary condition. Thus, one should get for any $\omega \in \Omega$,
\be \label{esscont}
\lim_{\Delta \uparrow \mathbb{Z}^2} \sup_{\omega^1,\omega^2 \in \Omega}  \Big| \mu \big[f |\mathcal{F}_{\Lambda^c} \big](\omega_\Delta \omega^1_{\Delta^c}) - \mu \big[f |\mathcal{F}_{\Lambda^c} \big](\omega_\Delta\omega^2_{\Delta^c})\Big|=0
\ee
which yields an (almost-sure) asymptotically  weak dependence on the conditioning, which can be seen as an extended  Markov property. 
In particular, for Gibbs measures the conditional probabilities always have continuous versions, or, equivalently,
 there is no point of essential discontinuity, in the following sense: 

\begin{definition}[Essential discontinuity]\label{essdiscdef}
A configuration $\omega \in \Omega$ is said to be a point of essential  discontinuity for a conditional probability of $\mu \in \mathcal{M}_1^+$ if there exists $\Lambda_0 \in \s$, $f$ local, $\delta >0$, such that for all $\Lambda$ with $\Lambda_0 \subset \Lambda$ there exist    $\mathcal{N}_\Lambda^1(\omega)$ and $\mathcal{N}_\Lambda^2(\omega)$, two open (or at least positive-measure) neighborhoods of $\omega$, such that
$$
\forall \omega^1 \in \mathcal{N}_\Lambda^1(\omega),\; \forall \omega^2 \in \mathcal{N}_\Lambda^2(\omega), \;\Big| \mu \big[f |\mathcal{F}_{\Lambda^c} \big](\omega^1) - \mu \big[f |\mathcal{F}_{\Lambda^c} \big](\omega^2)\Big| > \delta
$$
or equivalently
\be\label{essdisc}
\lim_{\Delta \uparrow \mathbb{Z}} \sup_{\omega^1,\omega^2 \in \Omega}  \Big| \mu \big[f |\mathcal{F}_{\Lambda^c} \big](\omega_\Delta \omega^1_{\Delta^c}) - \mu \big[f |\mathcal{F}_{\Lambda^c} \big](\omega_\Delta\omega^2_{\Delta^c})\Big|  > \delta.
\ee
\end{definition}
In the generalized Gibbsian framework, one also says that such a configuration is a  {\em bad configuration} for the considered measure, see {\em e.g.\ }\cite{ALN08}. In virtue of the Kozlov-Sullivan characterization of Gibbs measures,   the existence of such bad configurations is characteristic of a non-Gibbsianness of the associated measures. 

In this paper, we pursue the detection side of the Dobrushin Program of restoration of Gibbsianness, and extend previous non-Gibbs results obtained in the context of Decimations, the simplest RG transformations so far. We extend the Israel-van Enter-Fernandez-Sokal original approach (which was worked out in most detail for the $2d$-$n.n.$ Ising model) in two directions : to higher-dimensional long-range Ising models, and to,  possibly anisotropic,  rotator models.

{\bf {\em Decimated Measures :}}

Denote by $\mu^+$ a particular Gibbs measure, for the Ising or rotator models described below. We shall consider  some plus phase, obtained  as the weak limit (\ref{muplusminus}) with plus boundary conditions, but there is no need to be more  precise  for the moment (see next Section). We shall thus submit these  Gibbs measures to the {\em decimation transformation}  :
\be \label{DefDec}
 T \colon (\Omega,\mathcal{F})  \longrightarrow (\Omega',\mathcal{F}')=(\Omega,\mathcal{F}); \; 
\omega \; \;   \longmapsto \omega'=(\omega'_i)_{i \in
\mathbb{Z}^2}, \; {\rm with} \;  \omega'_{i}=\omega_{2i}
\ee
Denote by $\nu^+\coloneqq T \mu^+$ the decimated measure, formally defined as an image measure {\em via}
$$
\forall A' \in \mathcal{F'},\; \nu^+(A')=\mu^+(T^{-1} A')=\mu^+(A) \; {\rm where} \; A=T^{-1} A'= \big\{\omega: \omega'=T (\omega) \in A' \big\}.
$$
We distinguish between original and image sets using  primed notation, although by  rescaling  the configuration spaces $\Omega$ (original) and $\Omega'$ (image) are identical. 

{\bf For Ising spins}, the original measure $\mu^+$ we consider will be the standard plus phase obtained by taking the homogeneous all plus b.c.\ ${\bf +}$, defined by ${\bf +}_i=+1$ for all $i \in \Z^2$.

To investigate potential points of essential discontinuity for the image measure, we will choose, inspired by the magnetization being the order parameter,   the local function $f(\sigma')=\sigma'_{(0,0)}$, and will need to be able to evaluate
\be \label{condmagn}
 \nu^+[\sigma'_{(0,0)}| \mathcal{F}_{\{(0,0)\}^c} ](\omega') = \mu^+[\sigma_{(0,0)} | \mathcal{F}_{S^c} ](\omega),\; \nu^+-{{\rm a.s.}},
\ee
where $S^c=(2 \mathbb{Z}^2) \cap \{(0,0)\}^c= (2 \mathbb{Z}^2)^c \cup \{(0,0)\}$ is not  finite: {\em the conditioning  is {\bf not} on the complement of a finite set}; so that DLR Equations (\ref{DLR1}) do not hold directly for such set $\Lambda=S$. 

 In this Ising-spin ferromagnetic context, one can extend the formalism and use the Global Specifications derived for them in \cite{FP97}, to get a family  $\Gamma^+$ of conditional probabilities such that $\mu^+ \in \mathcal{G}(\Gamma^+)$, built in the following Theorem \ref{globspe}, where $S= (2 \mathbb{Z}^2)^c \cup \{(0,0)\}$ consisting of the {\em odd integers plus the origin}. 

\begin{theorem}\cite{FP97, ELN17, ALN13}\label{globspe}
Consider any of our ferromagnetic   Ising models on $\Z^2$ at inverse temperature $\beta >0$ with specification $\gamma^J$   with  couplings $J(i,j) \geq 0$ defined for any pair $\{i,j\} \in \Z^2$. In particular consider its  extremal Gibbs measures $\mu^+$ and $\mu^-$ obtained by the weak limits (\ref{WLimits}).

 Define $\Gamma^+=(\Gamma_S^+)_{S \subset \Z^2}$ to be the family of probability kernels on $(\Omega, \mathcal{F})$  as follows:
\begin{itemize}
\item For $S=\Lambda$ finite, for all $\omega \in \Omega$,
$
\Gamma^+_\Lambda(d \sigma | \omega) \coloneqq \gamma^J_\Lambda (d \sigma |  \omega).
$
\item For $S$ infinite, for all $\omega \in \Omega$,
$
\Gamma^+_S(d\sigma | \omega)\coloneqq\mu_S^{+,\omega} \otimes \delta_{\omega_{S^c}}(d \sigma)
$
where the constrained measure $\mu_S^{+,\omega}$  is the weak limit obtained by  freezing in $+_S \omega_{S^c}$ on $\Lambda^c$: $$\mu_S^{+,\omega}(d \sigma_S)\coloneqq \lim_{\Delta \uparrow S} \gamma^J_\Delta (d \sigma\mid +_S \omega_{S^c}).$$
\end{itemize}
Then $\Gamma^+$ is a global specification such that $\mu^+ \in \mathcal{G}(\Gamma^+)$. It is moreover monotonicity-preserving and right-continuous, but not quasilocal when $\beta > \tilde{\beta}_c >0$. Similarly, one defines a monotonicity-preserving and left-continuous global specification $\Gamma^-$ such that $\mu^- \in \mathcal{G}(\Gamma^-)$.
\end{theorem}

Now, for any special  configuration\footnote{It will be  alternating  in our   non-Gibbs results, but  in principle  here it can be any configuration $\omega' \in \Omega'$.} $\omega'_{\rm alt}$, (\ref{condmagn}) reduces for $\nu^+$-a.e. $\omega' \in \mathcal{N}_{\Lambda'}(\omega'_{\rm alt}) $ to
\be \label{condmagn2}
 \nu^+[\sigma'_{(0,0)}| \mathcal{F}_{\{(0,0)\}^c} ](\omega') = \Gamma_{S}^+ [\sigma_{(0,0)} | \omega] \; \; \mu^+{\rm -a.e.} (\omega)
\ee
with $S =(2 \mathbb{Z}^2)^c \cup \{(0,0)\}$ and $\omega \in T^{-1} \{\omega'\}$ is  defined to coincide with the alternating configuration on the even lattice $2\Z^2$. The expression of the latter is provided in terms of the constrained measure $\mu^{+,\omega}_{(2\mathbb{Z}^2)^c \cup \{0\}}$, with $\omega \in T^{-1} \{\omega'\}$  so that we get for any $\omega' \in \mathcal{N}_{\Lambda'}(\omega'_{\rm alt})$,
$$
\nu^+[\sigma'_{(0,0)}| \mathcal{F}_{\{(0,0)\}^c} ](\omega) = \mu^{+,\omega}_{(2\mathbb{Z}^2)^c \cup \{(0,0)\}} \otimes \delta_{\omega_{2\mathbb{Z}^2 \cap \{(0,0)\}^c}} [\sigma_{(0,0)}].
$$
By monotonicity-preservation, it is explicitly built as the weak limit   obtained by  plus boundary conditions fixed after a freezing of $\omega$ on the even sites : $\forall \omega' \in \mathcal{N}_{\Lambda'}(\omega'_{\rm alt}), \forall \omega \in T^{-1} \{\omega'\},\;   $
\be \label{constrLimit}
\mu^{+,\omega}_{(2\mathbb{Z}^2)^c \cup \{(0,0)\}} (\cdot) =\lim_{\Delta \in\s,\Delta \uparrow (2 \mathbb{Z}^2)^c  \cup \{(0,0)\}} \gamma^J_\Delta (\cdot\mid +_{(2 \mathbb{Z}^2)^c  \cup \{(0,0)\})} \omega_{2 \mathbb{Z}^2 \cap\{(0,0)\}^c}).
\ee
Note that it is enough to consider this limit on a sequence of cubes  $\Delta_N=([-N,+N] \cap \mathbb{Z})^2$ in the original space. 

{\bf In the case of rotator spins}, the original measure $\mu^+$ we consider will be  the  Gibbs measure  obtained by taking the homogeneous b.c. $\nicevec{\sigma}^+$ with { vertical  } angle $\theta=\theta_i=+ \frac{\pi}{2}$ for any $i \in \Z^2$ where the latter denotes the angle between the vectorial configuration at site $i$ with the horizontal axis, $(\nicevec{\sigma}^+_i,\nicevec{e_1})=+  \frac{\pi}{2}$. The configuration itself is sometimes written shortly ${\bf \theta}$ or ${\bf + \frac{\pi}{2}}$ (in boldface). The limit for such homogeneous b.c. is known to exist from \cite{Pfis81} (while one can learn in \cite{FrPf83} that all the extremal Gibbs measures are obtained in that manner, with an homogeneous b.c. ${\bf \theta}$, with $\theta \in ]-\pi,+\pi]$).

Moreover, in this context with continuous symmetry, in two dimensions, the magnetization is always zero for short-range models by the famous Mermin-Wagner ``ban''. {The Mermin-Wagner theorem claims that in short-range models in dimensions 1 and 2,  continuous symmetries cannot be spontaneously broken.  Nevertheless, in the case of long-range slowly decaying potentials, {first investigated by Kunz and Pfister \cite{KP76}, non-zero magnetizations do  occur. Also, even if Mermin-Wagner applies, in some contexts, see {\em e.g.}\  \cite{ES2002}}, other order parameters can exist, leading to other manifestations of phase transitions. {Moreover, differently  from rotational long-range order, in superficially similar but non-rotationally-invariant models,  with random or periodic external fields,  {\em e.g.}\ longitudinal or vertical,  long-range order due ``spin-flop''  transitions etc. can occur,} so that Mermin-Wagner in fact doesn't provide as strict a ban on the occurrence of phase transitions as one might initially  imagine.} { In the course of our proofs,   various choices of local functions  $f$   can be made so that we investigate essential discontinuity by the evaluation of the conditional probabilities }
\be \label{condmagn-rota}
 \nu^+[f(\nicevec{\sigma}')| \mathcal{F}_{\{(0,0)\}^c} ](\nicevec{\omega}') = \mu^+[f(\nicevec{\sigma}')| \mathcal{F}_{S^c} ](\nicevec{\omega}),\; \nu^+-{{\rm a.s.}}
\ee
where $S^c=(2 \mathbb{Z}^2) \cap \{(0,0)\}^c= (2 \mathbb{Z}^2)^c \cup \{(0,0)\}$ is not  finite: {\em the conditioning  is {\bf not} on the complement of a finite set}; so that DLR Equations (\ref{DLR1}) do not hold. 

\subsection{Global Specification for Rotator Spins}

 Although Global Specifications, useful in such a situations, have  been proven to exist only in a more limited set-up in the study of ferromagnetic Ising  cases by Fern\'andez {\em et al.} in \cite{FP97}, we describe {now} how they can be extended thanks to the attractivity\footnote{For  extension or non-extension, depending on the dimension  to rotator or classical Heisenberg models of classical correlation inequalities (GHS  or Lebowitz inequalities),  see  Ginibre \cite{Gin70} or Lebowitz \cite{Leb74}.   See also Dunlop {\em et al.}\cite{Dun76, DunNew75}, Monroe \cite{Monroe75}, Ellis {\em et al.} \cite{EMN76}, Kunz {\em et al.}, \cite{KPV75, KPV76}, Romerio {\em et al.} \cite{RV74, RV75}.} of $XY$-models, $O(N)$-models with $N=2$, such  that the following analogue of Theorem \ref{globspe} is valid:

{\begin{theorem}\label{globspe-rotator}{\em [Global specification for $2d$-rotator spins]}
Consider any of our ferromagnetic rotator models on $\Z^2$ at inverse temperature $\beta >0$ with specification $\gamma^J$   with  (ferromagnetic) couplings $J(i,j)$ defined for any pair $\{i,j\} \in \Z^2$, and in particular its  extremal Gibbs measures $\mu^+$ and $\mu^-$,  respectively obtained by weak limits from the opposed angle-b.c.  ${\bf \theta^+} \equiv {\bf +\frac{\pi}{2}}$ or ${\bf \theta^-} \equiv{\bf -\frac{\pi}{2}}$.

 Define $\Gamma^+=(\Gamma_S^+)_{S \subset \Z^2}$ to be the family of probability kernels on $(\Omega, \mathcal{F})$  as follows:
\begin{itemize}
\item For $S=\Lambda$ finite, for all $\nicevec{\omega} \in \Omega$,
$
\Gamma^+_\Lambda(d \nicevec{\sigma} | \nicevec{\omega}) \coloneqq  \gamma^J_\Lambda (d \nicevec{\sigma} |  \nicevec{\omega}).
$
\item For $S$ infinite, for all $\nicevec{\omega} \in \Omega$,
\be \label{GammaRota}
\Gamma^+_S(d\nicevec{\sigma} | \nicevec{\omega})\coloneqq \mu_S^{+,\nicevec{\omega}} \otimes \delta_{\nicevec{\omega}_{S^c}}(d\nicevec{ \sigma})
\ee
where the constrained measure $\mu_S^{+,\nicevec{\omega}}$  is the weak limit obtained with freezing in $\nicevec{{\bf +}}_S \nicevec{\omega}_{S^c}$ on $\Lambda^c$: 

$$\mu_S^{+,\nicevec{\omega}}(d \nicevec{\sigma}_S)\coloneqq \lim_{\Delta \uparrow S} \gamma^J_\Delta (d \nicevec{\sigma}\mid \nicevec{{\bf +}}_S \nicevec{\omega}_{S^c}).$$
\end{itemize}
Then $\Gamma^+$ is a global specification such that $\mu^+ \in \mathcal{G}(\Gamma^+)$.  Similarly, one defines a monotonicity-preserving and left-continuous global specification $\Gamma^-$ such that $\mu^- \in \mathcal{G}(\Gamma^-)$.
\end{theorem}}

{{\bf Proof :}
{Let us describe how the extension of {the construction of a global specification in the attractive case \cite{FP97}}} {-- initially restricted to Ising state spaces $E=\{-1,+1\}$ -- extends to our continuous spin state-space $E=\mathbb{S}^1$ thanks to {our choice of} partial order $\leq_{\sin}$. Basically, we need to check that :

\begin{itemize}
\item The fact the initial order on configurations is only partial does not affect the use of the monotonicity-preserving property.

\item The fact that the state space becomes continuous does not affect measurability properties of the kernels defining the global specification, nor the use of monotone-convergence theorem.
\end{itemize}}

As in the Ising case, we define the kernels $\Gamma^+_S$ in two ways, depending whether $S$ is finite or infinite. We shall afterwards
 extend Lemma 3.1  and Lemma 3.2 from \cite{FP97} to our rotator landscape. 

First, let the local specification $\gamma^{J}$ and the global specification $\Gamma^{+}$ coincide on finite sets:
$$
\Gamma^{+}_{\Lambda} \equiv \gamma^{J}_{\Lambda}, \qquad \forall \Lambda \in S.
$$

Second, for any infinite set $S \subset \mathbb{Z}^{2}$ and any finite set $\Lambda$, by consistency we need
\be \label{key}
\Gamma^{+}_{S} = \Gamma^{+}_{S} \Gamma^{+}_{\Lambda} =  \Gamma^{+}_{S} \gamma^J_{\Lambda}.
\ee
Hence, for any function $f \in \mathcal{F}_{\rm loc}$ and any finite set $\Lambda$, it holds
$$
\Gamma^{+}_{S}(f \mid \nicevec{\omega}) = \int_{\Omega} \gamma^{J}_{\Lambda}(f \mid \nicevec{\sigma}) \Gamma^{+}_{S}(d\nicevec{\sigma} \mid \nicevec{\omega}) \underset{\rm properness}{=} \int_{\Omega} \gamma^{J}_{\Lambda}(f \mid \nicevec{\sigma}_{S}\nicevec{\omega}_{S^{c}}) \Gamma^{+}_{S}(d\nicevec{\sigma} \mid \nicevec{\omega}),
$$
{and this has to be valid for any choice of the boundary condition $\nicevec{\omega}$. Thus the infinite-volume probability measure $\Gamma^{+}_{S}(\cdot \mid \nicevec{\omega})$ is consistent with the so-called constrained specification $\gamma^{S,\nicevec{\omega}}$ defined for any $\nicevec{\eta}$ by
$$
\gamma_{\Lambda}^{S,\nicevec{\omega}}(\cdot \mid \nicevec{\eta}) \coloneqq  \gamma^{J}_\Lambda (\cdot \mid \nicevec{\eta}_{S}\nicevec{\omega}_{S^{c}} )
$$
for a frozen configuration $\nicevec{\omega}$.

To recover a global specification consistent with $\mu^+$, we shall 
perform a weak limit with plus b.c.\ directly on this constrained specification  $\gamma^{S,\nicevec{\omega}}=(\gamma_{\Lambda}^{S,\nicevec{\omega}}, \; \Lambda \in \mathcal{S})$}.

{Here $\gamma_{\Lambda}^{S,\nicevec{\omega}}(\cdot \mid \nicevec{\eta})$ defines a probability measure on $\left(\Omega_{S},\mathcal{F}_{S} \right)$ and $\gamma_{\Lambda}^{S,\nicevec{\omega}}(F \mid \cdot)$ is $\mathcal{F}_{S \setminus \Lambda}$-measurable for any choice of $F \in \mathcal{F}_{S }$, so that we have only to check properness and consistency of the family of kernels $\Gamma^+$ as defined by (\ref{GammaRota}). Then for any configuration $\nicevec{\omega}$, the choice of a candidate for an element of $\mathcal{G}(\gamma^{S,\nicevec{\omega}})$ ({in which the elements of the specification have to be {consistent according to} (\ref{key})}) is made as follows: for a finite set $\Lambda \subset S$, the configuration at $\Lambda^{c}$ is fixed at the $\theta^{+}$ values ({\em i.e.} $+ \pi/2$) if sites do not belong to $S^c$; otherwise, the configuration is frozen onto $\nicevec{\omega}$.

Such a candidate measure in $\mathcal{G}(\gamma^{S,\nicevec{\omega}})$, called the constrained measure $\mu^{+,\nicevec{\omega}}_{S}$, is defined {\em via} the weak limit\footnote{{We stress that the weak limit in (\ref{eq.wl1235}) for obtaining the constrained measure $\mu^{+,\nicevec{\omega}}_{S}$ with $\theta^{+}$-boundary conditions on $S$ is performed {\em after} the freezing into $\nicevec{\omega}$ on the set $S^{c}$.}}
\be
\label{eq.wl1235}
\mu^{+,\nicevec{\omega}}_{S} (\cdot) \coloneqq  \lim_{\Delta \uparrow S}\gamma_{\Delta}^{J}(\cdot \mid \nicevec{+}_{S}\nicevec{\omega}_{S^{c}}),
\ee
and gives rise, for any infinite set $S\subset \mathbb{Z}^{2}$, to the kernels
$$
\Gamma_{S}^{+}(d\nicevec{\sigma}\mid \nicevec{\omega})\coloneqq \mu^{+,\nicevec{\omega}}_{S} (d\nicevec{\sigma}_{S})  \otimes \delta_{\nicevec{\omega}_{S^c}}(d\nicevec{ \sigma}_{S^{c}})
$$
which may be also written as
$$
\Gamma_{S}^{+}(d\nicevec{\sigma}\mid \nicevec{\omega}) = \lim_{\Delta \uparrow S}\gamma_{\Delta}^{J}(d\nicevec{\sigma} \mid \nicevec{+}_{S}\nicevec{\omega}_{S^{c}}),
$$
to give a probability measure on $(\Omega,\mathcal{F})$. Let us now show, following almost {\em verbatim} Lemma 3.2 in~\cite{FP97}, that DLR consistency holds for the global specification candidate $\Gamma^+$, namely that for two infinite subsets $D_{1}\subset D_{2}$, we have: 
\be
\label{eq.globconst}
\Gamma^{+}_{D_{2}}\Gamma^{+}_{D_{1}}= \Gamma^{+}_{D_{2}}.
\ee
For a set $\Lambda_{1}\subset D_{1}$, and $f_{1}$ a $\Lambda_{1}$-local function, (\ref{eq.globconst}) may be stated as
$$
\int \Gamma^{+}_{D_{2}}(d\nicevec{\eta} \mid \nicevec{\omega}) \Gamma^{+}_{D_{1}}(f_{1} \mid \nicevec{\eta})= \Gamma^{+}_{D_{2}}(f_{1} \mid \nicevec{\omega}).
$$
Recalling that $\Gamma_{S}^{+}(d\nicevec{\sigma}\mid \nicevec{\omega})\coloneqq \mu^{+,\nicevec{\omega}}_{S} (d\nicevec{\sigma}_{S})  \otimes \delta_{\nicevec{\omega}_{S^c}}(d\nicevec{ \sigma})$, we just have to prove
$$
\int \mu^{+,\nicevec{\omega}}_{D_{2}}(d\nicevec{\eta}) \Gamma^{+}_{D_{1}}(f_{1}\mid \nicevec{\eta}_{D_{2}} \nicevec{\omega}_{D_{2}^{c}}) = \int \mu^{+,\nicevec{\omega}}_{D_{2}}(d\nicevec{\eta}_{D_{2}})f_{1}(\nicevec{\eta}_{S_{2}}).
$$ 
In order to have the claim we only need to check that for two non-negative, increasing (in the sense of the partial order $\leq_{\rm sin}$) functions $f_{1},f_{2}$ which are respectively $\Lambda_{1}$-local and $\Lambda_{2}$-local, with $\Lambda_{1} \subset D_{1}$, $\Lambda_{2} \subset D_{2} \setminus D_{1}$,
$$
\mathbb{E}_{\mu^{+}}(f_{1}f_{2})=\mathbb{E}_{\mu^{+}}(\Gamma^{+}_{D_{1}}(f_{1}\mid \cdot) f_{2}).
$$
So that in order to get it we only have to check that the monotonicity argument passes through as in \cite{FP97}.
In order to do so, let us recall now our important choice of partial order : $$\nicevec{\omega} \leq_{\sin} \nicevec{\omega}' \iff \sin \theta_{i}\leq \sin \theta'_{i} \; {\rm for \; all \;} i  \in \mathbb{Z}^{2}.$$
{ A key} observation to properly {use} the monotonicity of the initial configuration for this ($2d$) rotator model is that  any configuration $\nicevec{\sigma}$
 can be dominated as 
 $$
\nicevec{\sigma} \leq_{\rm \sin} \nicevec{\eta}_{\Lambda}\nicevec{+}_{\Lambda^{c}}.
$$
{Then
we have, for any $\Lambda \subset D_{1}$, by the defining properties of probability kernels,
\be
\begin{split}
\Gamma^{+}_{D_{1}}(f_{1} \mid \nicevec{\eta})  \underset{\rm Kernel \; monotony}{\leq}  \Gamma^{+}_{D_{1}}\left(f_{1} (\nicevec{\sigma}_{\Lambda} \nicevec{+}_{\Lambda^{c}})\mid \nicevec{\eta}\right) & =  \Gamma^{+}_{D_{1}}\left(f_{1} (\nicevec{\sigma}_{D_{1} \cap \Lambda} \nicevec{+}_{D_{1}\cap \Lambda^{c}})\mid \nicevec{\eta}_{D^{c}_{1}}\right)\\
& =  \gamma^{J}_{\Lambda_{1}}\left(f_{1} (\nicevec{\sigma}_{\Lambda} \nicevec{+}_{D_{1}\cap \Lambda^{c}})\mid \nicevec{\eta}_{D^{c}_{1}}\right)\\
& =  \gamma^{J}_{\Lambda}\left(f_{1} (\nicevec{\sigma}_{\Lambda})\mid\nicevec{+}_{D_{1}\cap \Lambda^{c}} \nicevec{\eta}_{D^{c}_{1}}\right) \\
 &= \gamma^{J}_{\Lambda}\left(f_{1} (\nicevec{\sigma}_{\Lambda})\mid\nicevec{+}_{D_{1}} \nicevec{\eta}_{D^{c}_{1}}\right).
\end{split}
\ee
So that
\be
\label{ineq.AZE566}
\Gamma^{+}_{D_{1}}(f_{1} \mid \nicevec{\eta}) \leq \gamma^J_{\Lambda}\left(f_{1} \mid \nicevec{+}_{D_{1}} \nicevec{\eta}_{D^{c}_{1}}\right).
\ee
Following (3.23) of \cite{FP97}, and using the right-continuity of the function $\nicevec{\eta} \mapsto \gamma_{D_{1}}^{+}(f_{1}\mid \nicevec{\eta})$ and monotonicity, we have
\be
\begin{split}
\mathbb{E}_{\mu^{+}}(\Gamma^{+}_{D_{1}}(f_{1}\mid \cdot )f_{2}) & \leq \lim_{\Delta \uparrow D_{2}}\gamma^{J}_{\Delta}(\Gamma^{+}_{D_{1}}(f_{1}\mid \cdot) f_{2})\leq \gamma^{J}_{\Lambda_{2}}(\Gamma^{+}_{D_{1}}(f_{1}\mid \cdot) f_{2}\mid \nicevec{+})\\
& \leq \int  \gamma^{J}_{\Lambda_{2}} (d\nicevec{\eta}\mid \nicevec{+}) \gamma^J_{\Lambda}\left(f_{1} \mid \nicevec{+}_{D_{1}} \nicevec{\eta}_{D^{c}_{1}}\right) f_{2}(\nicevec{\eta}) 
\end{split}
\ee
where in the second inequality we have also used the fact that $\Lambda_{2}$ is fixed once and for all, and in the last inequality we have used (\ref{ineq.AZE566}). Choosing appropriately $\Lambda_{2}$ such that $\Lambda_{2}\cap D_{1}=\Lambda$, we just recognize that the last term is
\be
\int  \gamma^{J}_{\Lambda_{2}} (d\nicevec{\eta}\mid \nicevec{+}) \gamma^J_{\Lambda}\left(f_{1} \mid \nicevec{+}_{D_{1}} \nicevec{\eta}_{D^{c}_{1}}\right) f_{2}(\nicevec{\eta}) = \int  \gamma^{J}_{\Lambda_{2}} (d\nicevec{\eta}\mid \nicevec{+}) f_{1}(\nicevec{\eta})f_{2}(\nicevec{\eta}),
\ee
so that we finally have the inequality
$$
\mathbb{E}_{\mu^{+}}(\Gamma^{+}_{D_{1}}(f_{1}\mid \cdot )f_{2}) \leq \mathbb{E}_{\mu^{+}}(f_{1}f_{2}).
$$
We wish now to show the converse inequality. Now we can still use monotony, but for a subset $M \subset \Lambda_{2}\cap D_{1}$, 
\begin{eqnarray*}
\mathbb{E}_{\mu^{+}}\left( f_{1}f_{2}\right)  &\leq& \lim_{\Lambda_{2}}\int  \gamma^{J}_{\Lambda_{2}} (d\nicevec{\eta}\mid \nicevec{+}) f_{1}(\nicevec{\eta})f_{2}(\nicevec{\eta})= \lim_{\Lambda_{2}}\int  \gamma^{J}_{\Lambda_{2}} (d\nicevec{\eta}\mid \nicevec{+}) \gamma^J_M(f_1 | \nicevec{\eta})f_{2}(\nicevec{\eta})\\
& \leq & \lim_{\Lambda_{2}}\int  \gamma^{J}_{\Lambda_{2}} (d\nicevec{\eta}\mid \nicevec{+}) \gamma^J_M(f_1 | \nicevec{+}_{D_{1}} \nicevec{\eta}_{D^{c}_{1}})f_{2}(\nicevec{\eta}) = \int  \mu^+ (d\nicevec{\eta}) \gamma^J_M(f_1 | \nicevec{+}_{D_{1}}\nicevec{\eta}_{D^{c}_{1}})f_{2}(\nicevec{\eta}).
\end{eqnarray*}
Then by Beppo Levi's theorem we have that
$$
\mathbb{E}_{\mu^{+}}\left( f_{1}f_{2}\right) \leq \lim_{M  \uparrow D_{1}} \int \mu^{+}(d\nicevec{\eta})f_{2}(\nicevec{\eta}) \gamma_{M}(f_{1} \mid  \nicevec{+}_{D_{1}}\nicevec{\eta}_{D^{c}_{1}})=\mathbb{E}_{\mu^{+}} \left( f_{2} \Gamma^{+}_{D_{1}}(f_{1}\mid \cdot )\right)  
$$
and hence consistency on infinite sets, $\Gamma^{+}_{D_{1}}=\Gamma_{D_1}^{+} \Gamma_{D_{2}}^{+}$.\hfill $\diamond$
}}

\medskip

Now, for any special  configuration $\nicevec{\omega}'_{\rm spe}$, (\ref{condmagn-rota}) reduces for $\nu^+$-a.e. $\nicevec{\omega}' \in \mathcal{N}_{\Lambda',\epsilon}(\nicevec{\omega}'_{\rm spe}) $ to
\be \label{condmagn22}
 \nu^+[f(\nicevec{\sigma}')| \mathcal{F}_{\{(0,0)\}^c} ](\nicevec{\omega}') =  \Gamma_{S}^+ [f(\nicevec{\sigma}')| \nicevec{\omega}] \; \; \mu^+{\rm -a.e.} (\nicevec{\omega})
\ee
with $S =(2 \mathbb{Z}^2)^c \cup \{(0,0)\}$ and $\nicevec{\omega} \in T^{-1} \{\nicevec{\omega}'\}$ is  defined to coincide with the chosen special configuration on the even lattice $2\Z^2$. The expression of the latter is provided in terms of the constrained measure $\mu^{+,\nicevec{\omega}}_{(2\mathbb{Z}^2)^c \cup \{ (0,0)\}}$, with $\nicevec{\omega} \in T^{-1} \{\nicevec{\omega}'\}$  so that we get for any $\nicevec{\omega}' \in \mathcal{N}_{\Lambda'}(\nicevec{\omega}'_{\rm spe})$,

$$
 \nu^+[f(\nicevec{\sigma}') | \mathcal{F}_{\{(0,0)\}^c} ](\nicevec{\omega}') =  \mu^{+,\nicevec{\omega}}_{(2\mathbb{Z}^2)^c \cup \{(0,0)\}} \otimes \delta_{\nicevec{\omega}_{2\mathbb{Z}^2 \cap \{(0,0)\}^c}} [f(\nicevec{\sigma}')].
$$
It can be explicitly built as the monotone weak limit   obtained by ${\bf +\frac{\pi}{2}}$-b.c. fixed after a freezing of $\nicevec{\omega}$ on the even sites :  $\forall \nicevec{\omega}' \in \mathcal{N}_{\Lambda'}(\nicevec{\omega}'_{\rm alt}), \forall \nicevec{\omega}\in T^{-1} \{\nicevec{\omega}'\},\;  $
\be \label{constrLimit2}
 \mu^{+,\nicevec{\omega}}_{(2\mathbb{Z}^2)^c \cup \{(0,0)\}} (\cdot) =\lim_{\Delta \in\s,\Delta \uparrow (2 \mathbb{Z}^2)^c  \cup \{0,0)\}} \gamma^J_\Delta (\cdot\mid \nicevec{+}_{(2 \mathbb{Z}^2)^c  \cup \{0,0)\})} \nicevec{\omega}_{2 \mathbb{Z}^2 \cap\{0,0)\}^c}).
\ee

We shall use these constructions in Section 4, depending on the model, and  adjust the function $f$ to get a non zero essential difference between different sub-neighbourhoods.}
\

\section{Ising and Rotator models: Phase Transitions}

In this section we review some of the low-temperature results which are known about the models we will consider.  We will consider   pair interactions with formal Hamiltonian

$$
H_\Lambda(\sigma_\Lambda)=- \sum_{\{i,j\}} J(i,j) \;   \langle {\sigma}_i \cdot  {\sigma}_j \rangle
$$

where ``$\langle \ \cdot \ \rangle$'' is either the scalar product (for Ising spins) or {\em an} inner product (for vector spins), as described below. In both the following scalar and vectorial cases, when there is no ambiguity, we denote $\gamma^J$ the corresponding Gibbs specification at inverse temperature $\beta>0$. We will for convenience restrict ourselves to spatial dimension $2$, 
although most of our arguments will be extendible to higher-dimensions. The fact that we consider two-component spins {comes from} that in this case we have correlation inequalities to our disposal, which don't hold for higher-component spins, so although a number of statements might still be true, the proofs won't generalise as immediately. 

\subsection{Ising Models on $\Z^2$}

The configuration space $(\Omega,\mathcal{F},\rho)$ given by the products
\begin{displaymath}
\Omega=\{-1,+1\}^{\Z^2} \; , \;
\mathcal{F}= \mathcal{P}(\{-1,+1\})^{\otimes \Z^2} \;  , \;
 \rho=\Big( \frac{1}{2}\delta_{-1}+\frac{1}{2}\delta_{+1} \Big)^{\otimes \Z^2}\;,
\end{displaymath}
where $\delta_i$ is the Dirac measure on $i \in E=\{-1,+1\}$.

The interaction is $\Phi=(\Phi_A)_{A \in \mathcal{S}}$ defined by $\Phi_A \equiv 0$ if $A \neq \{i,j\}$ and 
\begin{displaymath}
\Phi_{\{i,j\}}( \sigma)= 
 -J(i,j) \; \sigma_i  \cdot \sigma_j 
\end{displaymath}
where $J : \Z^2 \times \Z^2 \longrightarrow R^+$ is a ferromagnetic coupling function ($J(i,j) \geq0$).

 In our ferromagnetic cases, Ising specifications are {\em monotonicity-preserving} 
(or {\em attractive}) in the sense that for all bounded increasing functions 
$f$, and  $\La \in \s$, the function $\gamma_\La^J f$ is 
increasing as a consequence of the FKG
property :  spins have a tendency to align \cite{FKG}. 
Using as boundary conditions the extremal (maximal ``$+$'' and minimal ``-'') elements of this
order $\leq$ already allows to define  extremal elements of $\mathcal{G}(\gamma^J)$. 
\begin{proposition}\label{Wlimit}\cite{FP97}
The weak limits
\be \label{muplusminus}
\mu^-(\cdot) \coloneqq  \lim_{\La \uparrow \mathbb{Z}^d} \gamma_\La^J (\cdot | -)\; \; {\rm and} \; \; \mu^+(\cdot) \coloneqq  \lim_{\La\uparrow \mathbb{Z}^d}  \gamma_\La^J (\cdot | +)
\ee
are well-defined, translation-invariant and extremal elements of $\mathcal{G}(\gamma^J)$. For any $f$ bounded increasing, any other measure $\mu \in \mathcal{G}(\gamma^J)$ satisfies
\be \label{stochdom}
\mu^-[f] \leq \mu[f] \leq \mu^+[f].
\ee
Moreover, $\mu^-$ and $\mu^+$  are respectively left-continuous and right-continuous.
\end{proposition}

 For the standard $2d$-$n.n.$ model, the existence of a critical temperature  has been  established by Peierls in 1936 \cite{Peie,Grif} and we state here the results we need through the following theorem  on the structure of the set $\mathcal{G}(\gamma^{J})$ of Gibbs measures for the corresponding Ising  specification $\gamma^{J}$\cite{Aiz,CV,Geo88,Hig}.

\begin{theorem}\label{PTIsing} Let $\gamma^{\beta J}$ be the specification (\ref{Gibbspe}) with  $2d$-$n.n.$ Ising potentials  at temperature $\beta^{-1}>0$. 
Then there exists a critical inverse temperature $0 < \beta_c < + \infty$ such that
\begin{itemize}
\item $\mathcal{G}(\gamma^{J})=\{\mu_\beta \}$ for all $\beta < \beta_c$.
\item $\mathcal{G}(\gamma^{J})=[\mu^-_\beta , \mu^+_\beta]$ for all  $\beta > \beta_c$ where the {\em extremal phases} $\mu^-_\beta \neq \mu^+_\beta$ can be selected {\em via} ``$-$'' or ``$+$'' boundary conditions: for all $f \in \mathcal{F}_{\rm{qloc}}$,
\be \label{WLimits}
\mu^-_\beta [f]\coloneqq \lim_{\Lambda \uparrow \mathcal{S}} \gamma_\Lambda^J [f \mid -] \; {\rm and} \; \mu^+_\beta [f]\coloneqq \lim_{\Lambda \uparrow \mathcal{S}} \gamma_\Lambda^J [f \mid +].
\ee
Moreover, the extremal phases have opposite {\em magnetizations} $$m^*(\beta)\coloneqq \mu^+_\beta [\sigma_0]=-\mu^-_\beta [\sigma_0]>0.$$
\end{itemize}
\end{theorem}


 {\bf {\em We focus here on the following long-range, possibly anisotropic in space, extensions of the $n.n.$ case and consider the following  Ising models I1-I2-I3 :}}

\begin{description}

 \item{\bf Model I1. (Very) long-range,  anisotropic in space,  uniaxial}
 ($d=2$, $\alpha_{1} > 1$) : $J \geq 0$
 $$
 J^{n.n.,\alpha_1}(i,j)\coloneqq J \cdot  \mathbf{1}_{|i -j |=1} \cdot \mathbf{1}_{|i_1-j_1|=0}+ J \cdot |i_1-j_1|^{-\alpha_1}  \cdot \mathbf{1}_{|i_2-j_2|=0} 
 $$

{\bf Results on $\mathcal{G}(\gamma^J)$ :} As in the $2d$-$n.n.$ model, the set $\mathcal{G}(\gamma^J)$ of Gibbs measures coincides with the interval $[\mu^-,\mu^+]$ for long-range ferromagnetic  models in $1d$, even in the phase transition region $\alpha \in (1,2)$. For higher-dimensional long-range models however, in the isotropic case, there could  exist non-translation invariant extremal Gibbs measures, similar to the so-called {\em Dobrushin states} for $3d-n.n.$ Ising models. Although this is not likely to happen  in long-range isotropic models with $\alpha >2$, and it has been excluded  for decays $\alpha >3$ by  the general arguments  due to Dobrushin/Shlosman in 1985 \cite{DS85}, it may happen that interface states exist  in the anisotropic bi-axial cases allowing very long-range interactions (axial decay between 1 and 2). Indeed, in such anisotropic cases, due to the possibility of phase transition for one-dimensional polynomially decaying pair potentials for very long range decays $\alpha \in (1,2)$, an extension of the roughening  proof techniques used in the  $3d-n.n.$ model, shows that there exist non-translation-invariant extremal Gibbs measures (similar to Dobrushin states in higher dimensions), see \cite{vB75, BLOP79, CELNR18, ALN20}. There thus also could be non-translation Gibbs measures amongst the mixed states. 

Although the non-Gibbsianness of all decimated  Gibbs measures does not always follow from the non-Gibbsianness of the translation-invariant extremal states $\mu^-$ or $\mu^+$, in our models this will in fact be the case. 

  \item{\bf Model I2.  Bi-axial, (Very) long-range, anisotropic in space }

  ($d=2$, $\alpha_1,\alpha_2 > 1$) : $J \geq 0$
 $$
 J^{\alpha_1,\alpha_2}(i,j)\coloneqq J \cdot  |i_2-j_2|^{-\alpha_2}  \cdot \mathbf{1}_{|i_1-j_1|=0}+ J \cdot |i_1-j_1|^{-\alpha_1}  \cdot \mathbf{1}_{|i_2-j_2|=0} 
 $$

{\bf Results on $\mathcal{G}(\gamma^J)$ :} Same remark as Model I1: When one of the decay powers along an axis, $\alpha_1$ or $\alpha_2$, is less than $2$, there could be a phase transition for the restriction of the interaction  to a single  axis, and the proof  mechanism of {\em e.g.}\ van Beijeren \cite{vB75} applies, leading to rigid interface states (extremal and non-translation-invariant). Nevertheless our proof on non-Gibbsianness applies to all Gibbs measures, as we show that there exist two open sets in any neighborhood of  the discontinuity point, conditioned on which  the two conditional expectations differ by more than some given small  constant. And open sets have positive measure for all Gibbs measures.

  \item{\bf Model I3. Long-range, Isotropic in space }  ($\alpha > d=2$) : $J \geq 0$, $ i,j \in \mathbb{Z}^d$,
 $$
J^{iso,\alpha}(i,j)\coloneqq  J \cdot |i-j|^{-\alpha}
 $$

{\bf Results on $\mathcal{G}(\gamma^J)$ :} The full convex structure has not been proven to coincide with the simplex $[\mu^-,\mu^+]$ (although this is probaly the case, as  suggested  by the fact that the absence of   Dobrushin states has been proven  in \cite{CELNR18}), and no non-translation-invariant states can exist at low T \cite{DS85} for $\alpha >3$. We shall focus on the decimations of the plus or minus measures, $\mu^-$ and $\mu^+$, obtained by the standard b.c.\ procedure in (\ref{WLimits}) but, as in model I2, our proof is nevertheless valid for any Gibbs measure, without using its extremal decomposition.
 \end{description}

\subsection{Rotator Models on $\Z^2$}

When changing the values of the Ising spins from $E=\{-1,+1\}$ to unitary vectors belonging to the $N$-dimensional sphere $E=\mathbb{S}^{N-1}$, keeping the same type of pair potentials, we say that one considers {\em $O(N)$-models}\footnote{{This is due to the fact that the finite- volume Hamiltonian is left unchanged by the simultaneous action of an element of the (compact, Lie) orthogonal group on all the microscopic vectors.}}, also called {\em Heisenberg models}. In this work, for vectorial cases, we only consider the case of the one-dimensional sphere $\mathbb{S}^1$ (so $N=2$), also called {\em rotator} --or {\em $XY$}-- {\em  models}. Although, similarly to $2d$-$n.n.$ Ising models, a phase transition is known to happen in $3d$ (see {\em e.g.}\ Friedli-Velenik \cite{FV16}, Chapter 10, and references therein), it is also well known that for such models continuous symmetries cannot be broken in two dimensions. One says that there is {\em no magnetic long-range order}, but\footnote{Other types of transitions are possible, related to fluctuations of magnetization, called quasilong-range order in $2d$ or, beyond this case, other Berezinskii-Kosterlitz-Thouless transitions, see \cite{FV16} or references here.} if we impose periodic or random external fields in the direction of one of the spin components, at low temperature
  a so-called  {\em spin-flop} transition is possible  \cite{Craw2011,ER2008}. 
 This can be used to provide a non-Gibbsianness result for  the low-temperature decimated  $2d$-$n.n.$ rotator model, even  in absence of a phase transition for this model, but we leave this to further studies. 

In this paper, in order to be able to make use of correlation inequalities for vector spins, we restrict ourselves to spin dimension two, and focus on two  rotator models for which a phase transition is possible (in our DLR-terminology) : Anisotropic rotator models (models  V1) and long-range ones (models V2). 

Also called {\bf Vector spins} (a particular case of  Heisenberg or $O(N)$-models), our $2d$ rotator models can be seen as the continuous counterpart of the Ising spins; the single-spin space is the unit sphere of $\mathbb{R}^{2}$, {\em i.e.} the circle $E=\mathbb{S}^{1}$; the {\em a priori} measure $\rho_0$ is the normalized Haar
measure on it and the lattice is $\Z^{2}$, while we identify the circle   with the interval of angles $]-\pi,+\pi]$. More formally, we consider the configuration space to be the measurable space $(\Omega,\mathcal{E}^{\otimes \Z^2},\rho_0^{\times \Z})$ where

\begin{displaymath}
\Omega=(\mathbb{S}^1)^{\Z^d} \; , \;
\mathcal{E}=\mathcal{B}(]-\pi,+\pi]) \;  , \;
 \rho_{0}=\lambda_{]-\pi,+\pi]} \;.
\end{displaymath}
Here $\lambda_X$ denotes the Haar measure (Lebesgue normalized) on $X$.

 The reason why we focus on these models only, with $N=2$, 
  is that in such cases we shall be able to extend the Ising case thanks to correlation inequalities adapted to this vectorial context, that do not occur in higher spin dimensions, 
   see \cite{Simon2020} or later in our proofs. This two-components spin framework allows also a proper stochastic order that permits us to extend the concept of Global Specification to this vectorial case, as we did in the previous section.

At $d=2$ in the $n.n.$  case, the $O(N)$-model  becomes the celebrated (classical) $XY$-model. Here absence of spontaneous magnetization (Mermin-Wagner theorem \cite{MW1966}, for a recent extension see {\em e.g.}\  \cite{ISV2002}), plus some extra properties,  gives a unique shift-invariant rotation-invariant Gibbs measure at all temperatures, due to a result of \cite{BFL77}. At large distances, the spin-spin two point correlation function is exponentially decaying at small $\beta$~ \cite{MS77} but only algebraically decaying at large $\beta$~\cite{FrSp81}, beyond  the Berezinskii-Kosterlitz-Thouless transition. It has been sometimes called an ``infinite-order'' phase transition, since the corresponding free energy is infinitely differentiable at the transition point (notice that other spin models with $O(N)$-symmetry may display a first-order phase transition with a  positive latent heat, see \cite{ES2005}). \\
\smallskip

\noindent
{\bf Side remark on $d=1$}: \\
Note that at $d=1$ with long-range interactions,  for quadratically decaying interactions, for  Ising spins  a Thouless effect occurs, namely, a discontinuous jump of the magnetization from zero to a strictly positive value as $\beta$ is increased~\cite{ACCN,SS81}. However, this does not occur for vector spins \cite{Sim81}. For slower decaying long-range vector models, with decay power $1 < \alpha < 2$,   the behaviour is as for Ising spins, though, low-T magnetisation with an ordinary critical point.  See   \cite{FrSp83}, \cite{DumGarTas},  \cite{Geo88}, \cite{FV16} and corresponding bibliographical notes for further details. 
\bigskip

 We shall start from a  plus phase $\mu^+$ obtained  by a weak limit of some arbitrary b.c., using results from  Ruelle \cite{Ruelle69} that the pure phases at low temperature -- {\em i.e.}\ the extremal translation invariant equilibirum states -- are obtained by taking as boundary condition $\nicevec{\sigma}_i \equiv \theta$ for all $i$ outside $\Lambda$ where $\theta \in ]-\pi,+\pi]$ for us. Note that Fr\"ohlich-Pfister  \cite{FrPf83} prove also that {all} the pure phases are constructed in this manner.

We recall the standard results of the $n.n.$ model with usual scalar product in the next section (namely the Mermin-Wagner Theorem on absence of continuous-symmetry-breaking  in $d=2$ \cite{Mermin67, MW1966,Shl77,Geo88,FV16} {\em vs.}\ phase transition in $3d$, but also the Kunz-Pfister result for (very) long-range interactions on existence of a phase transition at low $T$ \cite{KP76,FILS78}). 

In the rest of the paper, and as the main novelty in this Gibbs {\em vs.} non -Gibbs framework, we consider rotator models that will be either anisotropic $n.n.$ (Section \ref{anisotropic-vector}) or isotropic long-range (Section \ref{iso-lr-vector}). We also provide hints on how to  exhibit a non-Gibbsian decimated measure in absence of phase transition for the original model, for the classical    $2d$-$n.n.$ rotator (where we suspect the existence of bad configurations due to "spin-flop").

The potentials will be of the form
\be \label{XY}
\Phi_A(\sigma)=\left\{
\begin{array}{lll}
\; J(i,j) \;\cdot  \langle \nicevec{\sigma}_i \cdot  \nicevec{\sigma}_j \rangle \; \; & &\textrm{if} \; A=\{i,j\} \;\\
\; 0 \; & &\textrm{otherwise}
\end{array} \right.
\ee
where $\langle \ \cdot \ \rangle $ is {\em some} inner product in $\R^2$. More precisely, we shall investigate and prove non-Gibbsianness of the decimated measure at low temperatures in the following models V1-V2-V3:

In all cases (including Anisotropic), monotonicity-preservation allows to get the well-defined weak limits as in the Ising case :

\begin{proposition}\label{Wlimit2}\cite{FP97}
Consider the $XY$-models defined from (\ref{XY}) with a ferromagnetic coupling function $J(\cdot,\cdot)$, and the boundary condition $\theta=\theta^+={\bf + \pi/2}$ and $\theta=\theta^-={\bf - \pi/2} $. Then the weak limits
\be \label{muplusminus2}
\mu^-(\cdot) \coloneqq   \lim_{\La \uparrow \mathbb{Z}^d} \gamma_\La^J (\cdot | \theta^-)\; \; {\rm and} \; \; \mu^+(\cdot) \coloneqq  \lim_{\La\uparrow \mathbb{Z}^d}  \gamma_\La^J (\cdot | \theta^+)
\ee
are well-defined, translation-invariant and extremal elements of $\mathcal{G}(\gamma^J)$. For any $f$ bounded increasing (for the {\em sin}-order defined previously), any other measure $\mu \in \mathcal{G}(\gamma^J)$ satisfies 
\be \label{stochdom2}
\mu^-[f] \leq \mu[f] \leq \mu^+[f].
\ee
Moreover, $\mu^-$ and $\mu^+$  are respectively left-continuous and right-continuous.
\end{proposition} 



Furthermore, it is possible to get an extremal decomposition in terms of these weak limits : Let ${\bf \theta}$ be an everywhere $\sigma_i=\theta \in ]-\pi,+\pi] $ configuration. Then for any ${\bf \theta}$, there exist  an extremal state $\mu^\theta$, so that any  $\mu \in \mathcal{G}(\gamma)$ is written
$$
\mu = \int_0^1 \alpha_\theta(\mu) d \nu_\theta.
$$

\begin{description}
 \item{\bf Model V1. Anisotropic spin interaction,  $n.n.$ planar rotator model}
 
This is a well-known model in theoretical physics, whose phase transition
 has been conjectured by Fisher in 1967~\cite{Fisher67}, and first proved by Bortz and Griffiths in 1972 \cite{BG72}  for large anisotropy, while Malyshev extended it for any anisotropy parameter $\kappa$ in \cite{Malyshev75}.   See also Costa-M\'ol \cite{CM2013}, Fr\"ohlich-Lieb \cite{FL78}, Hohenberg \cite{H67}, Kunz-Pfister-Vuillermot \cite{KPV75}, Romerio-Vuillermot \cite{RV74,RV75} for more complete studies. 

For a more precise discussion on the temperature-dependent anisotropy needed, see \cite{BF82}, for a relation to Renormalization Group see \cite{BSK81} (Remark 2, p.\ 426). See also \cite{BF81} for correlation inequalities needed in the course of our proofs.

We use the description and results of  Georgii (\cite{Geo88}, chapters 16-20 and bibliographical notes) and Friedli-Velenik (\cite{FV16}, Chapter 10). A parameter $\kappa \in (0,1)$  is introduced so that the ordinary inner product in $\R^{2}$ is substituted by 
$$
\langle \nicevec{\sigma}_{i} \cdot \nicevec{\sigma}_{j} \rangle_{\kappa} \equiv \sigma_{i_1}\sigma_{j_1}+\kappa \, \sigma_{i_2}\sigma_{j_2}, \; {\rm for} \; i=(i_1,i_2), j=(j_1,j_2) \in \Z^2.
$$ 
Thus we keep
the $n.n.$ coupling 
$
J^{n.n.}= J   \cdot \mathbf{1}_{|i - j |=1} 
 $ 
 and consider
  the pair potential $\Phi$ with $\Phi_A=0$ unless $ A=\{i,j\}, |i-j|=1$ where 
 $$
 \Phi_A(\nicevec{\sigma})= - J \langle \nicevec{\sigma}_{i} \cdot\nicevec{\sigma}_{j} \rangle_{\kappa}.
 $$

The parameter $\kappa \in [0,1]$ is here the anisotropy parameter (so that the $XY$-model is recovered at $\kappa=1$). One can learn in the references above or in detail in \cite{FV16}, Chapter 10,  that there are indeed two ground states for this system at $\kappa \in [0,1]$ (with corresponding  Gibbs measures $\mu^{+},\mu^{-} \in \mathcal{G}(\beta,\kappa)$, supposed to be the only extremal phases). Fluctuations of the finite-volume magnetization concentrate along the first direction at sufficiently low temperatures (Theorem 10.17). See also other investigations with external fields in \cite{Geo88}, p.\ 393, Chapter 18 (1st edition).

Consider the weak limits (they coincide with the preceding $ \mu^+,\mu^-$), proven to exist (see Fr\"ohlich/Pfister 1983 \cite{FrPf83}, Griffiths \cite{Griff71} or Friedli/Velenik \cite{FV16}, pp 412-413), for  $(\nicevec{e_1},\nicevec{e_2})$ being the canonical basis of $\Z^2$, 
\be \label{plusminusbc}
\mu^+ (\cdot) = \mu^{+\pi/2}(\cdot) : =\lim_{\Lambda \uparrow \mathcal{S}} \gamma^{J^\kappa} (\cdot | \nicevec{e_2}) \; {\rm and} \; 
\mu^+ (\cdot)= \mu^{-\pi/2}(\cdot)=\lim_{\Lambda \uparrow \mathcal{S}} \gamma^{J^\kappa} (\cdot | - \nicevec{e_2})
\ee

As we quote in Section 4, for an anisotropy parameter $0<\kappa<1$ phase transition is known to occur at low temperatures, with $\mu^+ \neq \mu^-$ in such a way that the vertical magnetizations differ, see (\ref{crucial}).

  \item {\bf Model V2. Isotropic long-range planar rotators} 
  
  It is the vectorial variant of Model I3, with the configuration space $\Omega=(\mathbb{S}^1)^{\Z^2}$ and the long-range pair potential given for ($\alpha > d=2$) with ferromagnetic couplings as follows : $J \geq 0$, $ i,j \in \mathbb{Z}^2$, and for any configuration $\nicevec{\sigma}$, 
 $$
J^{iso,\alpha}(i,j)\coloneqq  \frac{J}{|i-j|^{-\alpha}} \cdot \langle \nicevec{\sigma}_{i},\nicevec{\sigma}_{j} \rangle .
 $$
 
 Similarly to the other models, one can consider weak limits with our $+$ or $-$ boundary conditions, thanks to attractivity, in the same way as (\ref{plusminusbc}), and it has been known since Kunz-Pfister 1976 \cite{KP76}, using a domination of a vector-spin variant of Dyson's hierarchical models that there also spontaneous vertical magnetization (\ref{crucial}) holds. In this phenomenon, the extension of Griffiths'  inequalities, initiated by Ginibre \cite{Gin70,Gin71}, valid only for two-component spins,  is essential. 
\end{description}

\section{Decimation of $2d$ Long-Range Ising Models}

In the context of Ising spins, our extensions of the original Israel/van Enter-Fern\'andez-Sokal example of non-Gibbsianness deep in the phase transition region will be performed by adding anisotropic or isotropic polynomial (very) long-range terms in three steps, for models 1., 2. and 3.\ described below. In all models, we adapt the original proofs from  the $2d$-$n.n.$ and $1d$-Long-range contexts  \cite{EFS93,ALN13,ELN17}  to some  some long-range contexts in $2d$. 

We recall the strategy used there, involving similar alternating bad configurations $\omega'_{{\rm alt}}=(-1)^{i_1+i_2}$ for any site $i=(i_1,i_2)$, the same Global Specifications, but also slightly different energy estimates, also allowing  to use Equivalence of Boundary Conditions in order to shield off far away -direct- influences due to long-range terms.\\
{ We note that in the uniaxial or biaxial situation the total direct energy of a square of size $L$ with the outside of an annulus of size $N(L)$ is of order
$L^2{ [N(L)]}^{1- \alpha}$, while in the isotropic case it will be of order $L^2 {[N(L)]}^{2- \alpha}$. {This will give us a lower bound on the size of the annulus}. The indirect influence from across the boundary decays in unique phase in the annulus in a way which is uniform in the annulus size. Thus we expect that the annulus size needed in the proof will indeed be of the order suggested by the argument above.

Let us recall the results of \cite{BLP79} with our notations.  Two boundary conditions $\omega^1$ and $\omega^2$ are said to  have a {\em finite energy difference} when
$$
C_{\omega^1,\omega^2} : = \sup_\Lambda \sup_{\bar{\sigma}_\Lambda} \Big| \sum_{X \cap \Lambda^c \neq \emptyset, X \cap \Lambda \neq \emptyset} \Phi_X(\bar{\sigma}_\Lambda \omega^1_{\Lambda^c}) - \Phi_X(\bar{\sigma}_\Lambda \omega^2_{\Lambda^c}) \Big| < + \infty .
$$

In such a case, one concludes (\cite{BLP79}) that the infinite-volume limits with these boundary conditions will have an equivalent (or the same) extremal decomposition (extremal equivalence) so that it's enough to consider any of them (and forget any influences outside a large enough annulus). 

In fact, we should show that it is enough to estimate for the sum under the  absolute value the difference
$$
H_\Lambda^{+,\omega^1}(\sigma_\Lambda) - H_\Lambda^{+,\omega^1}(\sigma_\Lambda) 
$$

for relevant choices of b.c.\ in subsets  of neighborhoods where in  annuli we have suitably chosen sizes, where $H_\Lambda^{+,\omega}$ is the Hamiltonian for $\omega$-b.c. for the constrained (infinite-volume) Gibbs measures $\mu_S^{+,\omega}$ given by (\ref{constrLimit2}) and  (\ref{constrLimit}), living on  the set $S$ of internal spins (the spins living on a decorated lattice) plus the origin. 
 
In all  cases we consider,  as a necessary step we will need to prove the following bounded-energy estimate, where $\omega'^{+,1/2}$ will act as boundary conditions from the even spins to the internal spins, with $+$-b.c.\ very far away (that we will forget about, because asymptotically the reasoning will  obviously be justified)  :
\begin{lemma}\label{GeneralEstimate}
Write $\Lambda'=\Lambda'(L)=([-L,+L] \cap \Z)^2$ and $\Delta'=\Delta'(N)=([-N,+N] \cap \Z)^2$, with $N >L$. Then, there exist sub-neighborhoods $\mathcal{N}_1^+\coloneqq \mathcal{N}^+_{L,N}(\omega'_{{\rm alt}})$ and $\mathcal{N}_1^-\coloneqq \mathcal{N}^-_{L,N}(\omega'_{{\rm alt}})$ such that for $N=N(L)$ -- depending on the model and on the decay $\alpha$ --, and for  $\omega_1^+ \in \mathcal{N}_1^+$ and $\omega_2^+  \in \mathcal{N}_1^-$ b.c.\  we obtain {$\forall \Lambda \in \mathcal{S}$}
 
\be\label{bc}
\Big| H_{\Lambda,\omega_1^+}(\sigma_\Lambda) - H_{\Lambda,\omega_2^+}(\sigma_\Lambda) \Big| \leq C <  \infty.
\ee
\end{lemma}
 
In the following subsections, we describe the general common steps of the proofs, and then discuss in some detail the different adaptations to obtain some bounds on the size of the annulus (derivation of Lemma~\ref{bc} in both cases).

\subsection{Decimation of the Bi-axial 
$n.n.$/very Long-Range Cases $\alpha_1>1$}\label{Bi-Axial-VLR-2d-Ising}

We start as a warm-up with a mixed model, where the extension of the $n.n.$ case is performed only along vertical lines, where it is  very long-range with decay $\alpha_1>1$, in the ferromagnetic case $J \geq 0$ :
 $$
 J^{n.n.,1}(i,j)\coloneqq J \cdot  \mathbf{1}_{|i -j |=1} \cdot \mathbf{1}_{|i_1-j_1|=0}+ J \cdot |i_1-j_1|^{-\alpha_1}  \cdot \mathbf{1}_{|i_2-j_2|=0}.
 $$

The existence of a phase transition is deduced from the classical $n.n.$ case by stochastic domination, and we consider the decimation of the $+$ -phase $\mu^+$. As in this case the adaptation is similar to that of the $1d$ long-range case of \cite{ELN17}, 
 we only sketch it and focus on the main difference between this model and  the $n.n.$ $2d$-case : the choice of the size of the annulus $\Delta'_N=([-N,+N] \cap \Z)^2$   (and thus $\Delta_N=([-2N,+2N] \cap \Z)^2$)  so that the neighborhoods $\mathcal{N}_{L,N}(\omega'_{\rm alt})$ can exhibit different magnetization values.

Let us come back to the context of Section 1, where the decimation transformations have been defined. As already claimed, and proven by {\em e.g.}\ \cite{EFS93}, the special configuration, which will be shown to be a point of essential discontinuity, is as usual the neutral alternating configuration $\omega'_{\rm alt}$, defined naturally as
$$
(\omega'_{\rm alt})_i=(-1)^{(i_1+i_2)},\; \forall i \in  (i_1,i_2) \in \Z^2.
$$

The main results of this section, non-Gibbsianness at low temperatures, will as usual follow from the  observation that when a  phase transition holds for the original specification -- at low enough temperature   -- the same is true for the constrained specification {\em with alternating constraint}, albeit one needs even lower temperatures to have a phase transition, leading  to non-Gibbsianness of $\nu^+$. From this phase transition, one will get the following essential discontinuity result as soon as long-range effects are shielded off by choosing a large enough annulus -- the same as in \cite{ELN17} in this Subsection \ref{Bi-Axial-VLR-2d-Ising} --. The shielding-off of long-range effects will be (as in \cite{ELN17}) neglected by (\ref{bc}) and an argument similar to  Equivalence of boundary conditions as a {\em screening effect} will be  {combined with stochastic  domination by external  fields of homogeneous signs} to yield essential discontinuities {\em via} Lemma 2, and then non-Gibbsianness.

\begin{lemma}\label{keylemma}
Consider our Bi-axial $n.n.$/long range model with an horizontal range $1 < \alpha_1 \leq 2$, at sufficiently low temperature. Let $\Lambda' \subset \Delta' \in \mathcal{S}$ and consider two arbitrary configurations $\omega'^+ \in \mathcal{N}_{\Lambda',\Delta'}^+(\omega'_{\rm alt})$ and $\omega'^- \in \mathcal{N}_{\Lambda',\Delta'}^-(\omega'_{\rm alt}) $. Then  $\exists \delta >0$, and  $\exists \Lambda'_0$ big enough s.t.\ for some $\Delta' \supset \Lambda' \supset \Lambda'_0$ with $\Delta' \setminus \Lambda'$ chosen big enough compared to $\Lambda'$, 
for all $\omega^+ \in T^{-1} \{\omega'^+\}$ and all $\omega^- \in T^{-1} \{\omega'^-\} $ 
\be \label{keymagn}
\Big| \mu^{+,\omega^+}_{(2\mathbb{Z}^2)^c \cup \{0\}}[\sigma_0] -  \mu^{+,\omega^-}_{(2\mathbb{Z}^2)^c \cup \{0\}}[\sigma_0] \Big| > \delta.
\ee
\end{lemma}

{\bf Proof of Lemma \ref{keylemma} for anisotropic $2d$ $n.n.$/long-range:}

To prove non-Gibbsianness in our models, we need to prove an essential difference between the constrained magnetizations $$
M^+ = \mu^{+,\omega^+}_{(2\mathbb{Z}^2)^c \cup \{(0,0)\}}[\sigma_{(0,0)}] \; {\rm and} \; M^- = \mu^{+,\omega^-}_{(2\mathbb{Z}^2)^c \cup \{(0,0)\}}[\sigma_{(0,0)}]
$$
but to do so we first have to prove that the later are well-defined and independent of asymptotic effects.

Write $\Lambda'=\Lambda'(L)=([-L,+L]\cap \Z)^2$ and $\Delta'=\Delta'(N)=([-N,+N] \cap \Z)^2$, with $N >L$.

{Let us give some further, more precise, description of  our notations. In (\ref{keymagn}), the involved constrained measures $\mu^{+,\omega}_S$ are infinite-volume Gibbs measures on the internal spins defined in (\ref{constrLimit2}) by taking the weak limit with $+$-b.c.\ with the constraint that configurations coincide with $\omega$ on $S^c=(2\mathbb{Z}^2)^c \cup \{(0,0)\}$, {\em i.e.}\ the complement of the decorated lattice (plus the origin). 

{\bf Step 1 : Equivalence of b.c. to get rid of the long-range effects}.

Denote formally  by $H$ the Hamiltonian of the constrained specifications for $\omega_1^+$ and $\omega_2^+$ as prescribed. Proceeding as in \cite{ELN17},   one can bound uniformly in $L$ the relative Hamiltonians with either $\omega_1^+$ and $\omega_2^+$ b.c. to get (\ref{bc}) in this case, as soon as one takes $N=N(L)=O(L^{\frac{2}{\alpha -1}})$ (thus larger than in $1d$):
$$
\delta H_L^{+,\omega'_{1/2}}\coloneqq \Big| H_{\Lambda,\omega_1^+}(\sigma_\Lambda) - H_{\Lambda,\omega_2^+}(\sigma_\Lambda) \Big| \leq C <  \infty.
$$
 
To get it and (\ref{bc}) in this case, we use the long-range structure of the interaction to get a uniform bound
$$
C_{{\omega'^1,\omega'^2}}^{\Lambda,\Delta} = \delta H_L^{+,\omega'_{1/2}} \leq \sum_{x \in \Lambda_{2L}} 2 \sum_{k > 2N} \frac{1}{k^\alpha} < C,
$$
as soon as 
\be \label{annulusI}
N(L) >> L^{\frac{2}{\alpha-1}}
\ee
because then for any $\sigma_\Lambda \in \Omega_{\Lambda}$
$$
\delta H_L^{+,\omega'_{1/2}} \leq 2\cdot (2L)^2 \cdot \frac{(2N)^{1-\alpha}}{1-\alpha}=8 L^2 \cdot \frac{(2N)^{1-\alpha}}{1-\alpha}.
$$
We have $C=1$ when
$$
(2N)^{1-\alpha} = \frac{1-\alpha}{8} L^{-2}
$$
{\em i.e.}
$$
N=\frac{1}{2} \left( \frac{1-\alpha}{8} \right)^{1/1-\alpha} L^{2/(\alpha-1)}
$$
so that choosing $N=N(L)$ as  in (\ref{annulusI}) will do the
job.

Then, by \cite{BLP79} (see also \cite{FV16}),   all of the limiting Gibbs states obtained by these boundary conditions have the same measure-zero sets, and therefore an equivalent decomposition into extremal Gibbs states. The latter decomposition will in fact be presumably trivial here, as we shall see that the Gibbs measure will be unique, but this is not needed at this step. Thus one gets the same magnetisation on the different sub-neighborhoods : $M^+=M^+(\omega, N, L)= M^+(\omega_1^+, N, L)=M^+(\omega_2^+, N, L)$ is indeed independent of $\omega$ as soon as it belongs to the pre-image of the $+$-neighborhood of the alternating configuration.

{\bf Step 2 : Domination by uniform fields -- Uniqueness for invisible spins}

{Consider boxes $\Lambda'_{L}\subset \Delta'_{N}$ of linear dimensions $2L, 2N$ (see Fig.~\ref{fig.vlx451c}).
\begin{figure}[!hbtp]
\centering
\includegraphics[width=.75\linewidth]{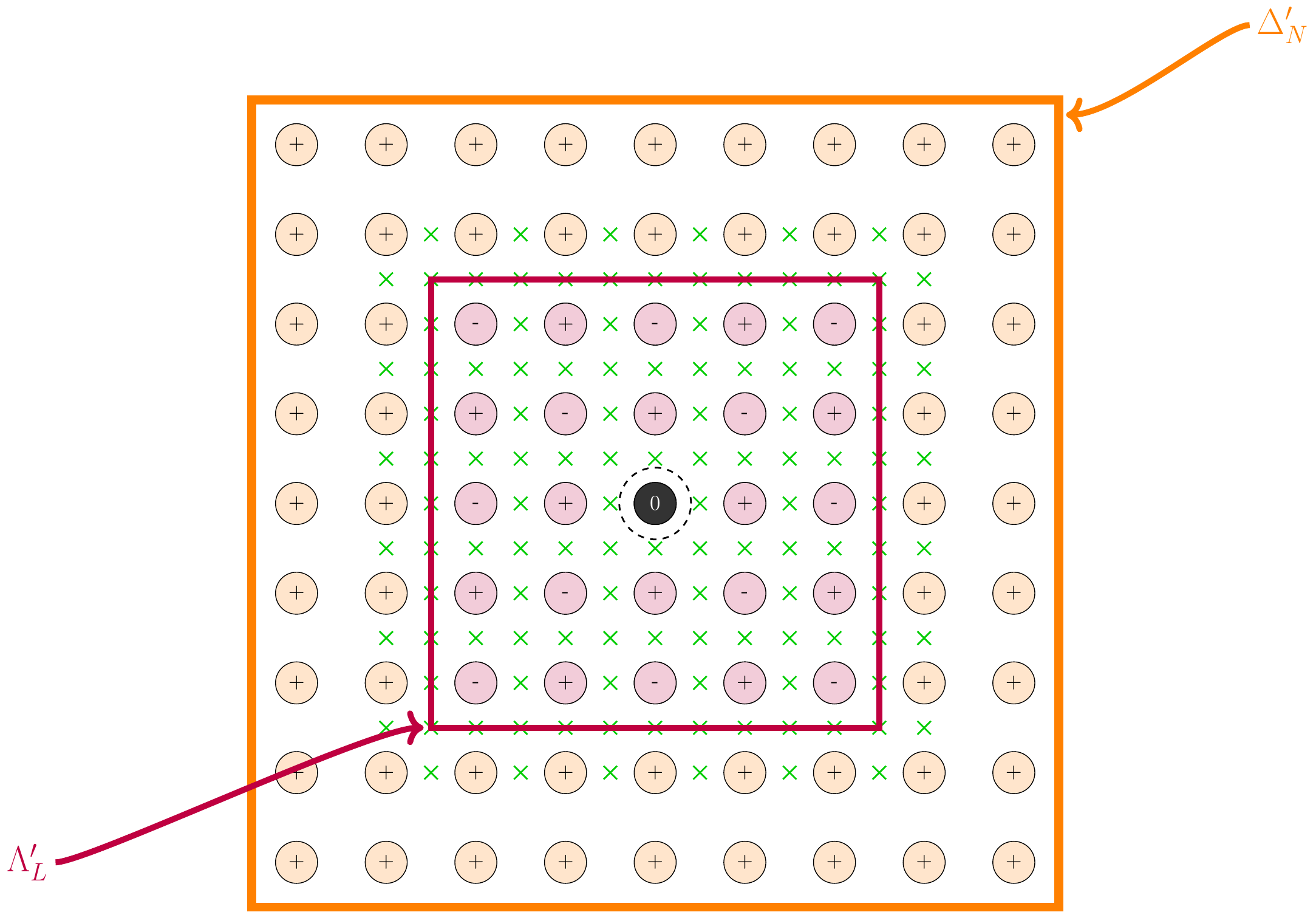}
\caption{{  The origin is in black and only the decorated lattice is shown (a portion of the original lattice is visible inside the box $\Lambda'_{L}$, depicted by green crosses).}}\label{fig.vlx451c}
\end{figure}

For any configuration $\omega'^+ \in \mathcal{N}_{\Lambda',\Delta'}^+(\omega'_{\rm alt})$, we have
\be \label{eq.k1243ptg}
\begin{split}
\nu^+[\sigma'_{(0,0)} | \mathcal{F}_{\{{(0,0)}\}^c} ] (\omega'^+)&= \mu^+[\sigma_{(0,0)} | \mathcal{F}_S^c ](\omega^{+}) \qquad \qquad \hfill \forall \omega^{+} \in  T^{-1}\{\omega'^+\}\\
&\underset{{\rm by}\; \eqref{constrLimit}}{=} \Gamma^{+}_{S}(\sigma_{(0,0)} \mid \omega^{+}) = \lim_{\Delta \uparrow S} \gamma_{\Delta}^{J}(\sigma_{(0,0)}\mid +_{S} \omega^{+}_{{S}^{c}})\\
&= \left < \sigma_{(0,0)}\right >^{h_j^{+}}_{S}.
\end{split}
\ee
for a box $\Delta$ satisfying $\Delta \subset (2 \Z^{2})^{c} \cup \{{(0,0)}\}$, where $\left < \cdots \right >^{h_j^{+}}_{S}$ denotes expectation with respect to the system at infinite volume $S$ with respect to the inhomogeneous effective field $(h_{j}^{+})_{j}$ depending on the lattice site $j$. The crucial point is that $\left < \sigma_{(0,0)} \right >^{h_j^{+}}_{S}$ in (\ref{eq.k1243ptg}) can be dominated below by an adaption of a rigorous argument due to Ruelle (see~\cite{Ruelle72}, Theorem 2). Indeed, observe that 
\be
h_{j}^{+} = h_{j}(\omega^{+})=\sum_{k \neq j} \frac{\omega^+_{k}}{k^{\alpha}} > \sum_{k \neq j}\frac{\omega^{+,0,L}}{k^{\alpha}} = c_{\alpha} \frac{1}{L^{\alpha-1}} > 0
\ee
where $\omega^{+,0,L}$ is the configuration which is zero inside $\Lambda_{L}$, and $+$ outside, and $c$ is some constant depending on $\alpha$. Thus, for any fixed $L$, the fields $h_{j}^{+}$ are strictly positive, and we can always choose $L,N$ sufficiently large such that $\left < \sigma_{0} \right >^{h_{(0,0)}^{+}}_{S} > \frac{\delta}{2}$, for some $\delta$ possibly depending on $L$. Hence we still get the essential discontinuity in the $+$ case, and an analogous statement holds for the measure $\nu^{-}$ {\it mutatis mutandi}, namely, $$
\nu^-[\sigma'_{(0,0)} | \mathcal{F}_{\{{(0,0)}\}^c} ](\omega'^{-})<\left < \sigma_{(0,0)} \right >^{h_{{(0,0)}}^{-}}_{S}<\frac{\delta}{2} <0.$$
Thus we get the following
}

\begin{theorem} 
For any  $1<\alpha_1 \leq 2$, at low enough temperature
 $\beta > \beta_c^{J(\alpha_1,*)}$, 
the decimated measure  $\nu^+$ of the plus phase $\mu^+$ for the anisotropic long-range Ising model I1, $\nu^+=T \mu^+$, is non-quasilocal, hence non-Gibbs, and so are all decimated Gibbs measures.
\end{theorem}

\subsection{Decimation of the Bi-axial (very) Long-Range Model $\alpha_1,\alpha_2>1$}\label{2d-VLR-Anisotropic}

We focus on the very long-range cases where both $\alpha_1$ and $\alpha_2$ are between\footnote{It is not difficult to see that the other values are contained in the $n.n.$ cases (or direct Markov order-$R$ adaptations thereof). Note that it is in fact enough that one of the $\alpha$'s being between 1 and 2, because then a non-zero spontaneous magnetization would already occur.} $1$ and $2$.

Recall ($d=2$, $\alpha_1,\alpha_2 > 1$) : $J \geq 0$,
$$
 J^{(\alpha_1,\alpha_2)}(i,j)\coloneqq J \cdot  |i_2-j_2|^{-\alpha_2}  \cdot \mathbf{1}_{|i_1-j_1|=0}+ J \cdot |i_1-j_1|^{-\alpha_1}  \cdot \mathbf{1}_{|i_2-j_2|=0}.
 $$

To prove Lemma \ref{keylemma} in this anisotropic/long-range case, we need to adjust the size of the annulus by proceeding as in (\ref{annulusI}, with only slight differences in the computation of the $2d$-sums. Indeed, the energy difference estimation becomes, with the same notations
$$
C_{{\omega'^1,\omega'^2}}^{\Lambda,\Delta} = \delta H_L^{+,\omega'1/2} \leq \sum_{x \in \Lambda_{2L}} 2 \Big(\sum_{k_1 > 2N}  k_1^{-\alpha_1} + \sum_{k_2 > 2N}  k_2^{-\alpha_2} \Big).
$$
{Write $\alpha\coloneqq {\rm min}(\alpha_1,\alpha_2)$}. Then proceeding as in model I, we get  that for $N$ bigger than 
$$
N(L)=\frac{1}{2} \left( \frac{1-\alpha}{16} \right)^{1/1-\alpha} L^{2/(\alpha-1)}
$$
suffices. One can thus take the same annulus as in model I, {for $\alpha$ being the slowest decay between the vertical and the horizontal one.}

Proceeding as in the previous subsection, by using the global specification and dominations with external fields of  homogeneous signs, we eventually get 
 \begin{theorem} 
For any  $1<\alpha_1,\alpha_2 \leq 2$, at low enough temperature
 $\beta > \beta_c^{J(\alpha_1,\alpha_2)}$, 
the decimated measure  $\nu^+$ of the  plus phase $\mu^+$ for the anisotropic long-range Ising model I2, $\nu^+=T \mu^+$, is non-quasilocal, hence non-Gibbs, and so are all decimated Gibbs measures.
\end{theorem}

\subsection{Decimation of the Classical Isotropic Long-range Models with $\alpha >2$}
Recall our model in the $2d$-vectorial context : 
 ($\alpha > 2$) : $J \geq 0$, $ i,j \in \mathbb{Z}^d$,
 $$
J^{iso,\alpha}(i,j)\coloneqq  J \cdot |i-j|^{-\alpha}.
 $$

{\bf Proof of Lemma \ref{keylemma} for  isotropic  $2d$ long-range models:}

We proceed as in models I1 and I2 above, with a slightly different estimation of the energy difference due to the bi-dimensional character of the interaction, leading to double sums instead of single sums.

Indeed, by evaluating it, one gets, still with the same notations
$$
C_{{\omega'^1,\omega'^2}}^{\Lambda,\Delta}\leq \sum_{x \in \Lambda_{2L}} 2 \cdot \sum_{y \in \Lambda_{2N}^c} \frac{1}{|y|^\alpha}
$$
which is of the order
$$
C=2 \cdot (2L)^2 \frac{(2N)^{2-\alpha}}{(2-\alpha)}
$$
and eventually a slightly smaller annulus size
$$
N>>  \frac{1}{2} \left( \frac{2-\alpha}{8} \right)^{\frac{1}{2-\alpha}} \cdot  L^{ \frac{2}{ \alpha-2} }
$$
and, of course for $\alpha >2$
$$
N(L)=L^{\frac{2}{\alpha-2}}.
$$

We get, proceeding similarly with global specifications and stochastic dominations, 
  \begin{theorem} 
For any  $1<\alpha \leq 2$, at low enough temperature
 $\beta > \beta_c^{J(\alpha)}$, 
the decimated measure  $\nu^+$ of the plus-phase $\mu^+$ for the anisotropic long-range Ising model I3, $\nu^+=T \mu^+$, is non-quasilocal, hence non-Gibbs, and so are all decimated Gibbs measures.
\end{theorem}

\section{Decimation of $2d$ Rotator Models}

\subsection{Decimation of $2d$ $n.n.$ Rotator Models with Spin Anisotropy}\label{anisotropic-vector}

We consider the anisotropic nearest-neighbour spin-interaction  model VI with an anisotropy favouring   the direction of the particular configuration ${\bf \theta}={\bf +\frac{\pi}{2}}$, in order to make use of the monotonicity and partial order described in the previous section. 

For an anisotropy parameter $\kappa \in (0,1)$, for configurations ($\nicevec{\sigma}_i)_{i \in \Z^2} \in (\mathcal{S}^1)^{\Z^2}$, we consider the $n.n.$ pair potential $\Phi=\Phi^\kappa$ defined so that $\Phi_A=0$ unless $ A=\{i,j\}, |i-j|=1$ where
 $$
 \Phi_{\{i,j\}}(\nicevec{\sigma})= - J \langle \nicevec{\sigma}_{i} \cdot \nicevec{\sigma}_{j} \rangle_{\kappa} =- J \big( \sigma_{i}^{(2)}\sigma_{j}^{(2)}+\kappa \, \sigma_{i}^{(1)}\sigma_{j}^{(1)} \big)
$$ 
and $n.n.$ coupling $$J^{n.n.}= J   \cdot \mathbf{1}_{|i - j |=1} 
 $$
 
where we have written $\sigma_{i}^{(1)}$ and $\sigma_{i}^{(2)}$ for the  the coordinates of the ``spin'' $\nicevec {\sigma}_i$ in the canonical basis $(\nicevec{e_1}, \nicevec{e_2})$. Note that the anisotropy here is along the vertical direction, which is important for what follows (and consistent with the partial order introduced in Section 2). This indeed provides at low temperature two opposite extremal Gibbs measures, a ``plus''-phase $\mu^+$ globally oriented upwards, and its opposite ``minus'' phase $\mu^-$ globally oriented downwards. 

Thanks to the partial order, the specification $\gamma^{J^\kappa}$ is monotonicity-preserving and the above ``extremal'' phases can be selected by the up or down b.c.\ $\pm \frac{\pi}{2}$, and we consider the weak limits, proven to exist (see Fr\"ohlich/Pfister 1983 \cite{FrPf83}, Griffiths \cite{Griff71} or Friedli/Velenik \cite{FV16}, pp 412-413), for  $(\nicevec{e_1},\nicevec{e_2})$ being the canonical basis of $\Z^2$, 
$$
\mu^+ (\cdot) = \mu^{+\pi/2}(\cdot) : =\lim_{\Lambda \uparrow \mathcal{S}} \gamma^{J^\kappa} (\cdot | \nicevec{e_2}) \; {\rm and} \; 
\mu^- (\cdot)= \mu^{-\pi/2}(\cdot)=\lim_{\Lambda \uparrow \mathcal{S}} \gamma^{J^\kappa} (\cdot | - \nicevec{e_2})
$$

with $\{\mu^-,\mu^+\} \in \mathcal{G}(\gamma^{J^\kappa})$. Our model is $n.n.$ thus the specification is also quasilocal (and in particular continuous in any direction \cite{FP97, MRVM99}, so also right- or left-continuous). All together, by right continuity and extremality (in the sense of the stochastic order), this allows us to extend the concept of global specification to this vectorial context : there exists a global specification $\Gamma^+$ for $\mu^+$ expressed in terms of weak limits of the constrained specification as in Theorem \ref{constrLimit}.
 
Now we consider the decimation of the $+$-phase $\nu^+=T \mu^+$. We prove that it is  a non-Gibbsian measure by proceeding as in Section 4. 

To get it  one can use the existence of a spontaneous vertical magnetization, proven by Ginibre \cite{Gin70}, Bortz {\em et al.} \cite{BG72}, Fr\"ohlich/Lieb \cite{FL78} or Kunz  {\em et al.} \cite{KPV75,KPV76} or Malyshev \cite{Malyshev75}, with the choice of the local function
$$
f(\nicevec{\sigma}')= \nicevec{\sigma}'^{(2)}_{\{0,0\}}
$$
corresponding to the vertical component of the spin at the origin (vertical magnetization).


What is important for us here is that at low temperature, for some values of $\kappa$ ($\in (0,1)$), one can define the weak limits $\mu^+$ and $\mu^-$ such that a {\bf vertical} spontaneous magnetization holds :
\be \label{crucial}
m^-(\beta)\coloneqq \mathbb{E}_\mu^-[ \nicevec{\sigma}^{(2)}_{\{0,0\}}] < 0 < \mathbb{E}_\mu^-[ \nicevec{\sigma}^{(2)}_{\{0,0\}}] \coloneqq m^+(\beta).
\ee

This is performed using  correlation inequalities coined by Ginibre \cite{Gin70} in this vertical context  (see {\em e.g.}\ the Peierls arguments of Bortz {\em et al.}\ \cite{BG72}, Malyshev \cite{Malyshev75}, or inequalities from Dunlop {\em et al.}\ \cite{DunNew75, Dun76} or Kunz {\em et al.}\ \cite{KPV76}).

Similarly the standard case of the $2d$-$n.n.$ Ising model, 
this can be use to estimate the conditional expectations of the decimated measures and prove essential discontinuities, either by proving spontaneous magnetization on the decorated square lattice, either by dominating the later by the one above on the full lattice (as we did for Ising spins).
 


To prove essential discontinuity, we consider again the alternating configuration 
$$
\nicevec{\omega}'_{\rm alt}= (-1)^{i_1+i_2}*\nicevec{e_1}
$$ 
as a bad configuration and prove that for some small $\epsilon>0$, for large enough volumes $\Lambda=\Lambda_L$, $\Delta=\Delta_{L+1}$, there exist two open subsets in all 
sub-neighborhoods, $\mathcal{N}^+\coloneqq \mathcal{N}_{\Lambda,\Delta,\epsilon}^{+\frac{\pi}{2}}(\nicevec{\omega}'_{\rm alt})$  and $\mathcal{N}^-=\mathcal{N}_{\Lambda,\Delta,\epsilon}^{-\frac{\pi}{2}}(\nicevec{\omega}'_{\rm alt})$ such that
 for $L,N$ large enough, for all $\nicevec{\omega}'^+ \in \mathcal{N}^+$ and $\nicevec{\omega}'^- \in \mathcal{N}^-$, 
 
 \be \label{essdicvec}
 \Big| \nu^+\left[\nicevec{\sigma}'^{(2)}_{\{0,0\}} \mid \mathcal{F}'_{\{0,0\}^c}\right] (\nicevec{\omega}'^+)- \nu^+\left[\nicevec{\sigma}'^{(2)}_{\{0,0\}} \mid \mathcal{F}'_{\{0,0\}^c}\right] (\nicevec{\omega}'^-) \Big| > \delta
 \ee
 for some $\delta >0$.  {Note that in these open subsets, the spins in the volume of size $L$ are $\varepsilon$-close to the alternating configuration, in the annulus they are $\varepsilon$-close to either a plus or a minus configuration, and outside the annulus they are arbitrary.  The constant $\varepsilon$ can be chosen small enough so that the total energy differences from the pure alternating and plus or minus configurations remains small.}

 To get (\ref{essdicvec}), we now use the extension of  Global Specifications  -- originally stated in \cite{FP97} for Ising spins only ($E=\{-1,+1\}$ -- to this (compact) vectorial context (see Section 1). Write again $\Gamma^+$ for such a global specification consistent with $\mu^+$. The magnetizations of above reads, as in the Ising spin context, as :
 $$
\langle {\nicevec{\sigma}'}_{(0,0)}^{(2)} \rangle^+_{S,{\omega'}} = \nu^+\left[\nicevec{\sigma}'^{(2)}_{(0,0)} \mid \mathcal{F}'_{(0,0)^c}\right] ({\nicevec{\omega}'}^+)= \mu^+\left[\nicevec{\sigma}^{(2)}_{(0,0)} \mid \mathcal{F}'_{S^c}\right] ({\nicevec{\omega}'}^+))= \Gamma_S^+ \big(\nicevec{\sigma}^{(2)}_{(0,0)} | {\nicevec{\omega}'}^+\big)
 $$
 where $S$ is as in the Ising context, the complement of the even sites plus the origin, which is not the complement of a finite set : 
 $
 S^c = (2 \Z^2) \cap \{(0,0)\}^c.
 $
 
 Similarly on for $\nicevec{\omega}'^- \in \mathcal{N}^- $,
  $$
\langle {\nicevec{\sigma}'}_{(0,0)}^{(2)} \rangle^-_{S,{\omega'}} = \nu^+\left[\nicevec{\sigma}'^{(2)}_{\{0,0\}} \mid \mathcal{F}'_{\{0,0\}^c}\right] ({\nicevec{\omega}'}^-)= \mu^+\left[\nicevec{\sigma}^{(2)}_{\{0,0\}} \mid \mathcal{F}'_{S^c}\right] ({\nicevec{\omega}'}^-))= \Gamma_S^+ \big(\nicevec{\sigma}^{(2)}_{\{0,0\}} | {\nicevec{\omega}'}^-\big).
 $$

Now, using (\ref{crucial}), we get (\ref{essdicvec}) because for any $\delta >0$ we can take $L,N$ large enough so that there exists a positive external field $h$ with
$$
\langle {\nicevec{\sigma}'}_{(0,0)}^{(2)} \rangle^+_{S,{\omega'}} > \langle {\nicevec{\sigma}'}_{(0,0)}^{(2)} \rangle^h > \frac{\delta}{2}
$$
and this yields essential discontinuity and non-Gibbsianness at low temperature. $\hfill \diamond$
\subsection{Decimation of Planar Long-Range Rotator Models }\label{iso-lr-vector}

In the context described in Section 2, we proceed similarly to prove the non-Gibbsianness of the measures $\mu^{+\pi/2}$ and $\mu^{-\pi/2}$, for a 
bad configuration  chosen to be the configuration  $\nicevec{\sigma}^{\theta,{\rm alt}}$ where for $i=(i_{1},i_{2})$,

$$\nicevec{\sigma}^{\theta,{\rm alt}}_i = (-1)^{i_1 +i_2} \mathbf{\theta}_{i}.$$

The proof requires only to combine the  Equivalence of boundary conditions {\em à la} Bricmont-Lebowitz-Pfister \cite{BLP79,ELN17} used in  Model I3 of this paper (with exactly the same estimate), so that Lemma \ref{Bi-Axial-VLR-2d-Ising} holds, with the {\em same} choice of annulus, with the use of the  Global specification performed in the model V1 above. Using exactly the same techniques, we get an essential discontinuity and non-Gibbsianness of the decimated measure for this model. Just as before, we get open sets by having the local configurations in sufficiently small intervals, both in volume and annulus. 

On these open sets $\mathcal{N}^-$ (res. $\mathcal{N}^+$), one selects on the  invisible spins the negatively (resp.\ positively) magnetized phase obtained by Kunz-Pfister \cite{KP76}, where here the magnetization has to be understood in the sense of the local function ``vertical magnetization'', as in the previous anisotropic model 
$$
f:\nicevec{\omega}' \longmapsto f(\nicevec{\omega}') = {\omega}'^{(2)}_{(0,0)}.
$$

Then, the strictly negative (resp.\ positive) values coined by Kunz-Pfister for very long ranges $\alpha \in (2,4)$ yields a significant difference on the two neighborhooods of the alternating configuration, leading to non-Gibbsianness at low temperature after the use of global specifications (valid in this context):

\begin{theorem} 
For any  $2<\alpha \leq 4$, at low enough temperature
 $\beta > \beta_c^{J(\alpha)}$, 
the decimated  measure  $\nu^+$ of the plus-phase $\mu^+$ for the anisotropic long-range rotator model V2, $\nu^+=T \mu^+$, is non-quasilocal, hence non-Gibbs, and so are all decimated Gibbs measures.
\end{theorem}

\subsection{Decimation of Rotator Models, Extensions  to Different Dimensions.}

 In fact, this argument also works immediately for the one-dimensional case with $1 < \alpha < 2$.\\
 Moreover, in higher dimensional $n.n.$ models, as well as long-range models the analysis also applies.  We just have to notice that the conditioned models are again vector models which have phase transitions, proven either by invoking
 \cite{FSS76} or \cite{BalabanOC} for $n.n.$ models, and either one of these papers or \cite{KP76} combined with correlation inequalities,  for various long-range models. \\
 Concluding,  in all examples discussed, decimating a low-temperature Gibbs measure at sufficiently low temperatures results in a non-Gibbsian measure.

 \section{Comments and Perspectives}
 In an upcoming Part 2 we will discuss cases where there is a difference in analysing the phase transition structure of the original model and suitably conditioned (on bad configurations) models.   This will in particular include  a number of borderline cases, in which the original models have no rotation-symmetry breaking, but a properly chosen conditioning may induce a transition which breaks a discrete -- spin-flip-like -- symmetry, due to the occurrence of a ``spin-flop'' transition.  In contrast to what happens with stochastically evolved measures \cite{ER2009}, the alternating configuration in decimated measures will be a continuity point in these examples, but we can find different spin configurations which act as discontinuity points. These borderline cases occur  in dimensions $d=1$ and $d=2$: in particular are included in  dimension $d=2$ either short-range models or long-range  rotator models with decay at last as fast as $1/r^4$, and  in dimension $d=1$ the $1/r^2$-rotator model.

 {Our results about the non-Gibbsianness of decimated Gibbs measures  apply at very low temperatures; it is not to be expected that they extend all the way to (or even above) the critical temperature, in view  of the analysis of the quite similar Ising situations for $n.n.$ models in general dimensions  and long-range models in $d=1$, in \cite{HallerKen1996,Ken}. \\
  Note that  some of the models in the class we discuss above are not fully understood, even at the physics level. In particular, it has recently been suggested \cite{GDRT} that for some decay powers the long-range isotropic vector models may have an intermediate Kosterlitz-Thouless phase in between the high-temperature regime with summable decay and the low-temperature regime 
where spontaneous magnetisation occurs.  It would of course be of great interest if this could be proven, but this seems not within reach to us by the methods at our disposal.

{\bf Acknowledgments:} 
Research has been partially supported within the CNRS International Research Program Bézout-Eurandom ``Random Graph, Statistical Mechanics and Networks'', supported by Laboratory LAMA (CNRS UMR8050), Bézout federation (CNRS Unit FR3522), Labex Bézout (ANR-10-LABX-58) and Eurandom (TU/e Eindhoven). M.~D'A.\ is grateful to S.~Caracciolo for providing reference~\cite{GDRT}.

\end{document}